\documentclass[fleqn,usenatbib]{mnras}

\usepackage{newtxtext,newtxmath}
\usepackage[T1]{fontenc}
\usepackage{ae,aecompl}

\usepackage{graphicx}	% Including figure files
\usepackage{amsmath}	% Advanced maths commands
\usepackage{amssymb}	% Extra maths symbols

\title[Have we measured the $\lambda \le 500\ \mu$m background?]{Have we seen all the galaxies that comprise the cosmic infrared background at 250\,$\mu$m $\le \lambda \le$ 500\,$\mu$m?}

\author[S. Duivenvoorden et al.]{S. Duivenvoorden,$^{1}$\thanks{E-mail: S.Duivenvoorden@Sussex.ac.uk}
S. Oliver,$^{1}$
M. B{\'e}thermin,$^{2}$
D. L. Clements,$^{3}$  
G. De Zotti,$^{4}$ \newauthor  
A. Efstathiou,$^{5}$
D. Farrah,$^{6,7}$      
P. D. Hurley,$^{1}$
R. J. Ivison,$^{8,9}$
G. Lagache,$^2$
D. Scott,$^{10}$  \newauthor 
R. Shirley,$^{1,11,12}$ 
L. Wang,$^{13,14}$ 
M. Zemcov$^{15}$ \\
$^{1}$Astronomy Centre, Department of Physics and Astronomy, University of Sussex, Brighton BN1 9QH\\
$^{2}$Aix-Marseille Univ., CNRS, LAM, Laboratoire d'Astrophysique de Marseille, 13013 Marseille, France\\
$^{3}$Astrophysics Group, Imperial College, Blackett Laboratory, Prince Consort Road, London SW7 2AZ, UK\\
$^{4}$INAF-Osservatorio Astronomico di Padova, Vicolo dell Osservatorio 5, I-35122 Padova, Italy \\
$^{5}$School of Sciences, European University Cyprus, Diogenes Street, Engomi, 1516 Nicosia, Cyprus\\
$^{6}$Department of Physics and Astronomy, University of Hawaii, 2505 Correa Road, Honolulu, HI 96822, USA \\
$^{7}$Institute for Astronomy, 2680 Woodlawn Drive, University of Hawaii, Honolulu, HI 96822, USA \\
$^{8}$European Southern Observatory, Karl-Schwarzschild-Stra{\ss}e 2, D-85748 Garching, Germany\\
$^{9}$Institute for Astronomy, University of Edinburgh, Royal Observatory, Edinburgh EH9 3HJ, UK\\
$^{10}$Department of Physics and Astronomy, University of British Columbia, 6224 Agricultural Road, Vancouver, BC V6T-1Z1, Canada\\
$^{11}$Instituto de Astrof\'{i}sica de Canarias, E-38205 La Laguna, Tenerife, Spain\\
$^{12}$Universidad de La Laguna, Dpto. Astrof\'{i}sica, E-38206 La Laguna, Tenerife, Spain\\
$^{13}$SRON Netherlands Institute for Space Research, Landleven 12, NL-9747AD, Groningen, the Netherlands \\
$^{14}$Kapteyn Astronomical Institute, University of Groningen, Postbus 800,
NL-9700 AV, Groningen, the Netherlands\\
$^{15}$Center for Detectors, School of Physics and Astronomy, Rochester Institute of Technology, Rochester, NY 14623, USA\\
}
\date{Accepted 2019 October 29. Received 2019 October 7; in original form 2019 May 17}

\pubyear{2018}

\begin{document}
\label{firstpage}
\pagerange{\pageref{firstpage}--\pageref{lastpage}}
\maketitle

% Abstract of the paper
\begin{abstract}
The cosmic infrared background (CIB) provides a fundamental observational constraint on the star-formation history of galaxies over cosmic history. We estimate the contribution to the CIB from catalogued galaxies in the COSMOS field by using a novel map fitting technique on the \textit{Herschel} SPIRE maps. Prior galaxy positions are obtained using detections over a large range in wavelengths in the $K_{\rm s}$--3\,GHz range. Our method simultaneously fits the galaxies, the system foreground, and the leakage of flux from galaxies located in masked areas and corrects for an ``over-fitting" effect not previously accounted for in stacking methods. We explore the contribution to the CIB as a function of galaxy survey wavelength and depth. We find high contributions to the CIB with the deep $r$ ($m_{\rm AB} \le  26.5$), $K_{\rm s}$ ($m_{\rm AB} \le  24.0$) and 3.6\,$\mu$m ($m_{\rm AB} \le  25.5$) catalogues. We combine these three deep catalogues and find a total CIB contributions of 10.5 $\pm$ 1.6, 6.7 $\pm$ 1.5 and 3.1 $\pm$ 0.7\,nWm$^{-2}$sr$^{-1}$ at 250, 350 and 500\,$\mu$m, respectively. Our CIB estimates are consistent with recent phenomenological models, prior based SPIRE number counts and with (though more precise than) the diffuse total measured by FIRAS. Our results raise the interesting prospect that the CIB contribution at $\lambda \le 500\,\mu$m from known galaxies has converged. Future large-area surveys like those with the Large Synoptic Survey Telescope are therefore likely to resolve a substantial fraction of the population responsible for the CIB at 250\,$\mu$m  $\leq \lambda \leq$ 500\,$\mu$m.
\end{abstract}

% Select between one and six entries from the list of approved keywords.
% Don't make up new ones.

%%%%%%%%%%%%%%%%%%%%%%%%%%%%%%%%%%%%%%%%%%%%%%%%%%

%%%%%%%%%%%%%%%%% BODY OF PAPER %%%%%%%%%%%%%%%%%%

\section{Introduction}

The diffuse extragalactic cosmic infrared background \citep[CIB, e.g.,][]{1996A&A...308L...5P} is caused by the re-radiation of absorbed UV and optical light emitted by young stars and (for a small fraction) active galactic nuclei (AGN). This thermal re-radiation contributes approximately half of the radiation we receive from extragalactic sources \citep[e.g.][]{2001ARA&A..39..249H,2018arXiv180203694H}. It is therefore important to understand which sources are responsible for this CIB, as they are the likely contributors to the star-formation rate density of the Universe \citep[e.g.][]{2014ARA&A..52..415M}.

The aim of this paper is to measure the contribution of galaxies detected in different wavelength bands to the CIB. The result can be used as a practical indicator of  what depth of data is needed to detect a significant fraction of the star-forming populations that cause the CIB, which is part of the aim of future generation large area surveys like the Large Synoptic Survey Telescope \citep[LSST;][]{2008arXiv0805.2366I}. We can furthermore use the results to give new and more accurate lower limits for the {\em total} CIB. These more accurate limits can be used to constrain galaxy evolution models.

The Far Infrared Absolute Spectrophotometer (FIRAS) instrument aboard the Cosmic Background Explorer \citep[\textit{COBE};][]{1994ApJ...420..457F} was designed to measure the cosmic microwave background spectrum, but the data could also be used to measure other physical quantities, including the CIB \citep{1998ApJ...508..123F,1999AA...344..322L}. FIRAS was able to measure the total CIB due to the presence of a cold external calibrator, a facility that more recent space based telescopes like the \textit{Herschel Space Observatory} \citep{herschel} and \textit{Spitzer} \citep{2004ApJS..154....1W} lacked. Due to the absence of this absolute measurement and a high thermal foreground from the warm telescope, each of the \textit{Herschel} maps have the mean of the map subtracted, resulting in a map with a average signal of zero. To measure the total flux in confused maps we therefore need to find the sum of the flux density of the individual sources contributing to these confused maps \citep{2006A&A...451..417D}.

Relatively few extragalactic sources are directly detected with \textit{Herschel}, with the integrated flux density of those galaxies being a factor of about 7 lower than the total radiation received as the CIB \citep{2010A&A...518L..21O}. Recent work in deblending the \textit{Herschel} maps \citep{2016MNRAS.460..765W,2017MNRAS.464..885H} reveals that it is possible to assign the flux density in the confused \citep[e.g.][]{2010A&A...518L...5N} \textit{Herschel} SPIRE \citep{2010A&A...518L...3G} maps to sources that are detected in higher resolution optical/NIR images. The question now arises: what depth of data do we need to effectively deblend these images? 

To calculate new bounds for the CIB we will use a novel map fitting analysis based on \texttt{SIMSTACK} \citep{2013ApJ...779...32V} applied to the \textit{Herschel} SPIRE \citep{2010A&A...518L...3G} maps in the COSMOS field \citep{2007ApJS..172..150S}. This field contains very deep catalogues in various wavelength bands and is therefore ideal for creating deep prior lists. The $\sim$2\,deg$^2$ size of the COSMOS field is another advantage compared to other deep fields which tend to be $<$1\,deg$^2$. In the near future, large area surveys will obtain data with the COSMOS depths over areas $\gg$ 100 deg$^2$ which could be used to find the optical/NIR counterparts of dusty star-forming galaxies over larger areas of the sky observed by \textit{Herschel}.

The paper is structured as follows. In Section \ref{sec:data} we introduce the different sets of prior catalogues we use for our map fitting. In Section \ref{sec:met} we explain our map fitting method and we test our method in Section \ref{sec:test}. Our results are described in Section \ref{sec:res} and discussed in Section \ref{sec:dis}. Our conclusions can be found in Section \ref{sec:con}.

\section{Data} \label{sec:data}

\subsection{HELP database}

Most of the data described below is part of the \textit{Herschel} Extragalactic Legacy Project \cite[HELP\footnote{\url{http://hedam.lam.fr/HELP/}},][Oliver et al. in prep.]{2019MNRAS.tmp.2145S} database. HELP  aims to collate and homogenize observations from many astronomical observatories to provide an integrated data set covering a wide range of wavelengths from radio to UV. The key focus of the HELP project is the data from the extragalactic surveys from ESA's \textit{Herschel} mission, covering over 1300 deg$^2$. HELP will add value to these data in various ways, including providing selection functions and estimates of key physical parameters. The data set will enable users to probe the evolution of galaxies across cosmic time and is intended to be easily accessible for the astronomical community. The aim is to provide a census of the galaxy population in the distant Universe, along with their distribution throughout three-dimensional space.

\subsection{Prior catalogues} \label{sec:prior}

For the optical/NIR data sets we use the \cite{2016ApJS..224...24L} COSMOS2015 catalogue. From this catalogue we retrieve the $r$-band data, which were observed with the SUBARU Suprime-Cam as part of the COSMOS-20 project \citep{2007ApJS..172....9T,2015PASJ...67..104T}. The $r$-band data have a 3$\sigma$ depth of $m_{\rm AB}$ = 26.5 in a 3 arcsec aperture. We use the unflagged regions in the optical bands inside the COSMOS 2 deg$^2$ field, which leaves us with a total useful area of 1.77 deg$^2$ \citep{2016ApJS..224...24L}. We only select galaxies with a \texttt{SExtractor} flag of 3 or lower \citep{1996A&AS..117..393B}. With this flag we remove saturated or corrupted objects. We do keep neighbouring galaxies which could cause a potential bias (an effect we discuss below).

The VIRCAM instrument on the VISTA telescope was used to obtain the $K_{\rm s}$-band data as part of the UltraVISTA survey \citep{2012A&A...544A.156M}. Several ultra-deep stripes were observed, which covered a total area of 0.62 deg$^2$ \citep{2016ApJS..224...24L}. We will use both the deep and ultra-deep $K_s$ data, but we use the 3$\sigma$ depth of the deep data ($m_{\rm AB}$ = 24.0 in a 3 arcsec aperture) as a cut-off for the whole catalogue. The total area with deep or ultra-deep $K_{\rm s}$-band data covers 1.38 deg$^2$ inside the COSMOS 2 deg$^2$ field (excluding masked regions). 

%{\color{red} We can choose to use the ultra-deep areas only, but we will loose more than half of the area. I still need to test if there is a gain from using this smal}

IRAC channel-1 (3.6\,$\mu$m) observations consist of the first two-thirds of the SPLASH COSMOS data set, together with S-COSMOS \citep{S-cosmos} and smaller IRAC surveys in the COSMOS field (Capak et al. in prep.). The 3$\sigma$ depth cut-off for IRAC channel 1 is $m_{\rm AB}$ = 25.5, and the area covered is 1.77 deg$^2$ (excluding masked regions). This is the same area used for the $r$-band 

We use the COSMOS catalogues, as they contain deeper NIR and IR data from UltraVISTA and SPLASH than previous catalogues. The optical/NIR photometry is obtained using \texttt{SExtractor} dual-image mode, which is highly effective in finding and selecting galaxies. Due to the new data depth and the dual-image strategy, the galaxy samples are very complete \citep{2016ApJS..224...24L}, with a stellar mass limit for star-forming galaxies of $10^{10}$ M$_\odot$ at $z < 2.75$ and $10.8^{10}$ M$_\odot$ at $z < 4.8$. \cite{2015ApJ...809L..22V} used the \cite{2013ApJS..206....8M} catalogue to obtain the prior $K$-selected ($m_{\rm AB} = 23.4$) catalogue for stacking. However the \cite{2013ApJS..206....8M} catalogue only has 115\,000  galaxies within a 1.62 deg$^2$ area where the \cite{2016ApJS..224...24L} catalogue contains 149,000 galaxies with $m_{\rm AB} \leq 23.4$ over an area of 1.38 deg$^2$, and a total of 200,000 detected galaxies with $K_{\rm s} < 24.0$. We therefore expect that the percentage of the CIB we can resolve will be higher than that in \cite{2015ApJ...809L..22V}  due to the higher completeness.

In the mid-infrared we use the MIPS 24\,$\mu$m data obtained by \cite{2009ApJ...703..222L}. We select objects with $S_{24} >$ 80 $\mu$Jy ($m_{\rm AB} <$ 19.1) and that have a $3\sigma$ detection. The total area observed with the MIPS instrument is 2.27 deg$^2$.

The PACS \citep{pacs} 100\,$\mu$m data in COSMOS was observed as part of the PEP survey \citep{2011A&A...532A..90L}. The PACS catalogue contains 7443 sources with a 3$\sigma$ detection and $m_{\rm AB} \leq 14.8$, spanning an area of 2.1 deg$^2$. 

SPIRE \citep{2010A&A...518L...3G} data were obtained as part of the HerMES survey 4th data release \citep{hermes} and covers an area of 4.9 deg$^2$. We use the xID250 catalogues, which use the 250\,$\mu$m \texttt{starfinder} detections as prior information for the positions. We only select sources with a 5$\sigma$ detection above the instrumental noise. 

We use the SCUBA-2 (850\,$\mu$m) data observed as part of S2CLS \citep{2017MNRAS.465.1789G}. The catalogue produced by S2CLS contains 719 sources detected with a 3$\sigma$ detection within the 1.3 deg$^2$ observed with an RMS below 2\,mJy\,beam$^{-1}$.

The VLA 3\,GHz data \citep{2017A&A...602A...1S} covers an area of 3.1 deg$^2$, where a median rms of 2.3\,$\mu$Jy beam$^{-1}$ is reached in the central 2 deg$^2$ COSMOS area. We use 5.5$\sigma$ detected sources ($m_{\rm AB} \leq 21.4$) in the central 2\,deg$^2$ COSMOS area for our prior list. 

We furthermore test our method in different fields to obtain an estimate of the effect of cosmic variance. We picked the SERVS IRAC channel 1 catalogues \citep{2012PASP..124..714M} in the ELAIS-N1 and CDFS-SWIRE fields to perform this test. The depth of the SERVS catalogues is $m_{\rm AB} = 23.1$ and therefore two orders of magnitude shallower that the COSMOS SPLASH sample. For the ELAIS-N1 and CDFS-SWIRE fields we use the star masks provided by HELP to remove sources in our catalogue contaminated by stars or bright galaxies.

\subsection{Maps for fitting}

We use the SMAP \citep{2010MNRAS.409...83L,2013ApJ...772...77V} SPIRE maps described in \cite{2015ApJ...809L..22V} for our map fitting analysis. These maps have a pixel scale of 4 arcsec, which is smaller than the standard HerMES maps, which have a pixel scale of 6, 8.33 and 12 arcsec at 250, 350 and 500\,$\mu$m, respectively. The SPIRE maps are all mean-subtracted. 

We use the 250, 350 and 500 $\mu$m SPIRE maps in the COSMOS field for our main analysis and we use the maps in the ELAIS-N1 and CDFS-SWIRE fields to check our method against cosmic variance. For ELAIS-N1 and CDFS-SWIRE fields we use the nominal pixel size maps. Absolute calibration in SPIRE has a 5 per cent calibration uncertainty \citep{2010A&A...518L...3G}.

\subsection{Previous CIB estimates}

We have collated a number of previous estimates for the CIB to compare with our results (Table \ref{tab:CIB}). \cite{1998ApJ...508..123F} measured the the CIB from FIRAS measurements by removing foreground emission from interplanetary and Galactic interstellar dust. \cite{1999AA...344..322L} obtained a different estimate of the CIB with the same FIRAS measurements, which differ from each other by around 10 per cent, but are still consistent within error bars. The FIRAS-derived values are are dominated by systematics, where the main systematic uncertainty is the removal of the Galaxy. Higher resolution observations with \textit{Herschel} are not sensitive to this large scale Galactic emission.

Another method to measure the CIB is by adding (stacking) the flux density for all known galaxies in the Universe. This method can potentially miss a diffuse part of the CIB outside our own galaxy (if it exists). But the main problem with this method is that stacking in the highly confused SPIRE maps is non trivial (see Section \ref{sec:met}) and that it potentially misses the flux density of galaxies which are not detected. Therefore these measurements \citep{2009ApJ...707.1729M,2015ApJ...809L..22V} are technically a lower-limit of the total CIB. 

\cite{2015ApJ...809L..22V} used the earlier $K_{\rm s} < 23.4$ \cite{2013ApJS..206....8M} COSMOS catalogue as input for \texttt{SIMSTACK} to calculate the CIB at 250, 350, and 500\,$\mu$m. The maps were smoothed to a resolution of 300 arcsec to capture the contribution of faint (undetected) sources that are correlated, with the detected sources. In this work we will use deeper catalogues and we will fit simultaneously for the foreground. 

\cite{2016ApJ...827..108D} calculated deep galaxy number counts at the SPIRE wavelengths using $r$-band priors in the GAMA fields and $i$-band priors in the COSMOS field \citep{2016MNRAS.460..765W}. The obtained number counts where extrapolated to get the number counts for undetected galaxies.  The total values for the CIB obtained with this method are consistent with the FIRAS measurements.

The CIB can also be measured due to the effect of Lensing. This method looks at the deficit in background surface brightness in the central region of massive galaxy clusters \citep{2013ApJ...769L..31Z}. This measurement of the CIB does not include the galaxies which are part of, or in front of the clusters.

The CIB can also be calculated from the output of simulations. The Durham semi-analytic model \citep[\texttt{GALFORM};][]{2015MNRAS.446.1784C,2016MNRAS.462.3854L}, which realistically simulates clustering and optical magnitudes, was used to create a simulated catalogue. This optical catalogue is then used as input for the radiative transfer code to obtain $\lambda_{\rm rest} >70$\,$\mu$m flux density estimates. These values are slightly lower than (and at 250\,$\mu$m in rough 1$\sigma$ tension with) results from FIRAS.  On the other hand, the \cite{2017AA...607A..89B} simulation (which populates a dark matter lightcone with separately generated galaxies) and the \cite{2013ApJ...768...21C} simulation contains a higher flux density at the SPIRE wavelengths, which in both cases are in line with the FIRAS methods.

\begin{table}
	\begin{tabular}{l c c c} 
		Work & 250\,$\mu$m & 350\,$\mu$m & 500\,$\mu$m \\ \hline
		\cite{1998ApJ...508..123F}* & $10.3\pm3.2$ & $5.6\pm1.7$ & $2.4\pm0.7$\\
		\cite{1999AA...344..322L}* &$11.0\pm3.6$ & $6.2\pm2.0$ & $2.4\pm0.8$   \\
		\cite{2009ApJ...707.1729M}$\dagger$ & $8.60\pm0.59$ & $4.93\pm0.43$ & $2.27\pm0.20$ \\
		\cite{2013ApJ...769L..31Z}$^+$ & $8.3^{+1.4}_{-0.8}$\\
		\cite{2013ApJ...768...21C} & 12.4 & 7.9 & 3.7 \\
		\cite{2015ApJ...809L..22V}$\dagger$ & $9.82\pm0.78$ & $5.77\pm0.43$ & $2.32\pm0.19$ \\
		\cite{2016MNRAS.462.3854L} & 7.4 & 4.8 & 2.3 \\
		\cite{2016ApJ...827..108D}$\ddagger$ & $10.00\pm1.82$ & $5.83\pm1.17$ & $2.46\pm0.75$ \\
		\cite{2017AA...607A..89B} & 11.2 & 6.4 & 2.7
	\end{tabular}
	\caption{The total CIB in units of nW m$^{-2}$ sr$^{-1}$ at the SPIRE wavelengths as measured by FIRAS*, stacking$\dagger$, lensing$^+$, number counts$\ddagger$ and simulations.}
	\label{tab:CIB}
\end{table}

\section{Method} \label{sec:met}

We use an improved stacking analysis to measure the contribution to the CIB originating from galaxies detected in different catalogues. Stacking is equivalent to determining the covariance between a catalogue and a map \citep{2009ApJ...707.1729M}. In traditional stacking, a list of prior positions is used to add the map at those positions on the sky together. The noise in this ``stacked'' image will go down with $\sqrt{N}$, with $N$ the number of stacked positions. Normal stacking works well for confused data, as the mean contribution of the uncorrelated sources is zero. 

However, normal stacking can overestimate the flux density in maps which are clustered \textit{and} confused, as it will add the flux density from correlated sources to the galaxies in the stacking sample. To get around this problem \texttt{SIMSTACK} \citep{2013ApJ...779...32V} was developed, which measures this covariance between the map and a catalogue by simultaneously fitting all the known sources in the map.

Original \texttt{SIMSTACK} creates images with delta functions at the positions of galaxies in the prior catalogue. These images are  convolved with the instrument PSF. This results in a linear model for every pixel ($j$) in the map ($M$) with  the mean flux ($S_\alpha$) for galaxies in each list, $\alpha$, as a free parameter:
\begin{equation} \label{eq:point}
M^j = S_1C_{1}^j +...+ S_nC_{n}^j,
\end{equation}
where $C_{\alpha}^j$ is the beam-convolved, mean-subtracted image (this is the mean of the map, not a local mean) of the sources in list $\alpha$, at pixel $j$. This method should provide an unbiased estimate of the mean fluxes of the populations. 

There are, however, two problems with the traditional \texttt{SIMSTACK}, it does not fit the foreground nor does it consider signal arising from sources located in masked areas. These masked areas are regions on the sky where there are no observations for the prior catalogue or regions where these data are corrupted. The corrupted areas mainly arise due to the saturation of pixels by nearby bright galaxies or stars. No galaxies are detected in these masked areas, and therefore we should not use this area for our map fitting. 

Areas masked because they have not been observed or because of saturation due to bright stars should have a comparable value for the SPIRE intensity as non-masked areas. However, the masked regions provided by \cite{2016ApJS..224...24L} have a higher mean signal than non masked regions in the SPIRE map due to the presence of bright nearby sources. A naive application of {\sc simstack} on a mean-zero map would thus underestimate mean fluxes, even leading to negative flux densities.  

To solve these two problems we adjusted the \texttt{SIMSTACK} code to fit a foreground layer\footnote{We use foreground for the diffuse component, which consist of the emission from the telescope, Galactic emission and the emission from galaxies which are not correlated with the galaxies in our prior catalogues. This layer also incorporates the mean subtraction of the SPIRE maps.} ($F$) and leakage from flux from masked areas due to the large PSF ($S_L$) simultaneously with the pointing-matrix created in Equation \ref{eq:point}. As we are now fitting for a foreground there is no more need to mean subtract the beam convolved number of sources ($N_{\alpha}^j$) in a layer. The equation we are solving for the areas used in this work is therefore: 
\begin{equation} \label{eq:point2}
M^j = S_1N_{1}^j +...+ S_nN_{n}^j + S_LN_{L}^j + F,
\end{equation}
where the constant foreground, $F$, is not a function of pixel $j$.  

We recreate the SPIRE maps with holes on the positions of masked areas in the prior catalogue. We do not use our map fitter in those masked areas. Not using these areas is crucial, since otherwise flux from sources within those areas will be added to the foreground estimation. However, due to the large SPIRE beam there will still be excess flux from those masked areas within the fitted region. This excess flux would be added to the foreground layer (or to galaxies near the masked area), which causes an overestimate of the foreground over the whole field and therefore an underestimate of the prior galaxies flux densities. We solve this problem by adding the extra layer (Equation \ref{eq:point2}) to our fitting process, this being the convolution of the masked pixels with the SPIRE beam ($S_LN_{L}^j$). We provide a more detailed explanation when we describe the use of simulations in the next section.

\subsection{Tests on simulations} \label{sec:test}

We use the 2 deg$^2$ SIDES model simulation \citep{2017AA...607A..89B} to test our method. The SIDES simulation populates the halos in a dark-matter light cone with galaxies. For each galaxy a star-formation rate and hence spectral energy distribution is assigned and a gravitational lensing factor is calculated. From this simulation the observed flux densities are calculated between 24\,$\mu$m and 1.3\,mm. We create our own 4\,arcsec pixel SPIRE maps from the catalogue provided by \cite{2017AA...607A..89B}. We make these maps by smoothing the sources with a Gaussian PSF having a FWHM of 17.6, 23.9 and 35.2 arcsec for 250, 350 and 500\,$\mu$m, respectively. We then add Gaussian pixel noise with $\sigma=$ 5.7, 7.6, 13.4\,mJy, comparable to the values for the instrumental noise in the observations. These simulated SPIRE maps contain clustering, instrumental noise, and confusion noise, which makes them ideal to test our map fitting analysis. For the prior  lists we divide the sources into magnitude bins with a width of 0.4 magnitude, using the observed MIPS (24 $\mu$m) magnitudes.  We use all $10^6$ galaxies in the 2 deg$^2$ with 24\,$\mu$m magnitudes $<$ 26.4, these galaxies contribute  more than 99 per cent of the CIB in the SIDES model.

To test our map fitting algorithm we create the SPIRE maps from the \cite{2017AA...607A..89B} sources in several different ways:  including or excluding the effects of clustering; instrumental noise; and confusion. These variants test how our method performs in different simulations and predicts corrections for systematic effects. For the simulated maps we know that the foreground is zero. Due to the lack of a foreground we can test if our code works in the absence of this foreground layer; however, the real SPIRE maps will have a non-zero foreground and we therefore need to use the foreground layer for the real maps.  

\begin{figure*}
	\centering  % this centres figure in column
	\includegraphics[trim = 0mm 4mm 0mm 0mm,width = 2.0\columnwidth]{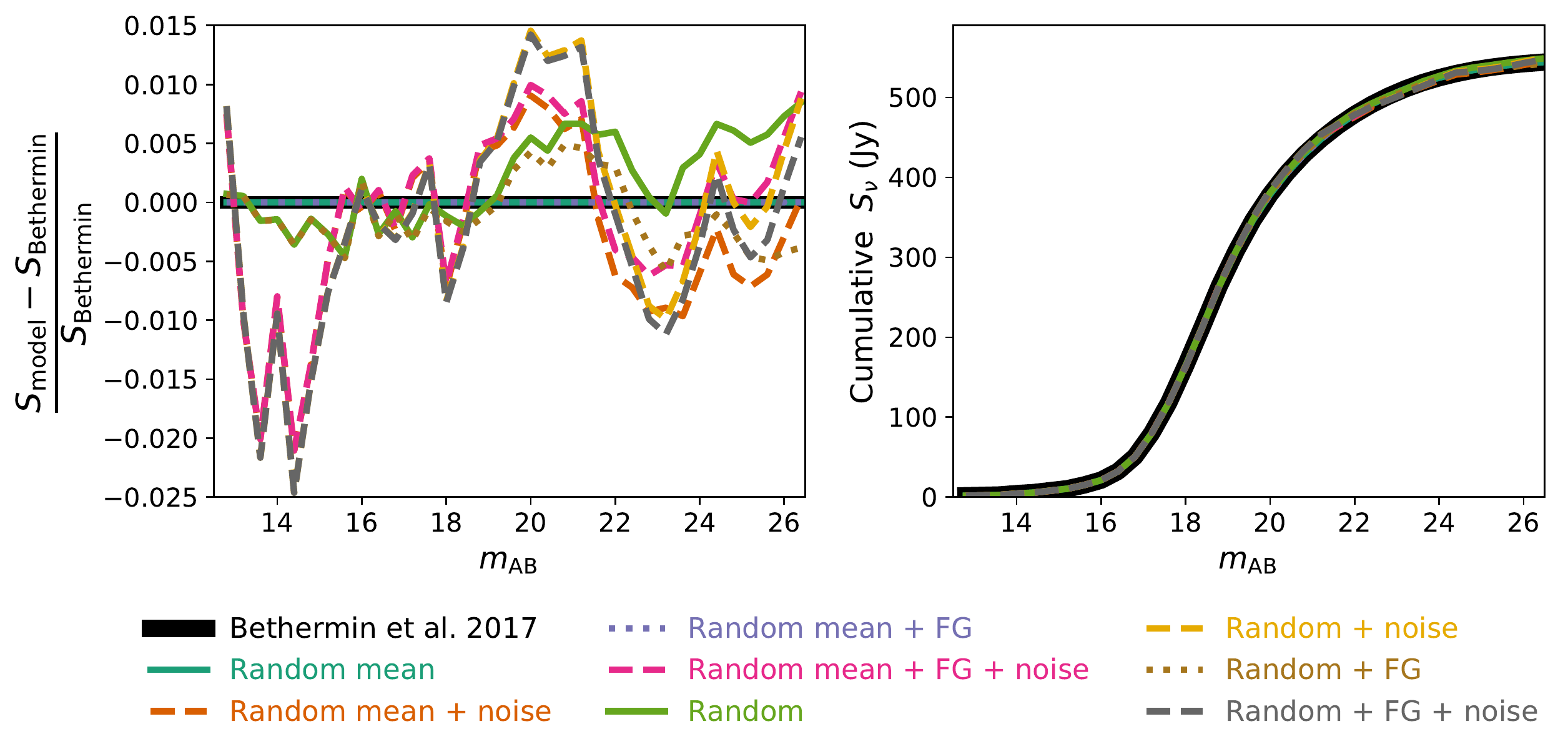}
	\caption{Testing our map fitting method at 250 $\mu$m for unclustered sources. In black is the ``truth" from the simulation. On the left we show the offset from the true answer and on the right the cumulative flux density as function of magnitude. In all tests the sources have random Poisson-distributed (uncorrelated) positions. Here ``Mean" indicates that the mean flux density of a population is used to create the map, ``noise" indicates that instrumental noise is added, and ``FG" indicates that we simultaneously fit for a foreground. For all models we are able to calculate the total CIB within 1 per cent accuracy.} 
	\label{fig:pois}
\end{figure*}

For the first series of tests we assign every source a random position in the map to avoid spatial correlations. In the first example we assign the mean flux density of the galaxies in a magnitude bin to every source within that bin. For this map our layer model (Equation \ref{eq:point2}) is perfect as our model is able to exactly describe the flux at every position in the map. Therefore, we obtain a $\chi^2 = 0$ without noise and a $\chi^2/N_{\rm pix} \sim 1$\footnote{The number of fitted parameter is orders of magnitude smaller than the number of pixels in the map, therefore the degrees of freedom $\approx N_{\rm pix}$ and the reduced $\chi^2 \approx \chi^2/N_{\rm pix}$.} when instrumental noise is included. These results are unaffected when we add a varying foreground to the test. The next tests are the same, but instead of the mean flux we use the actual flux density of each source. In this case we do not have a perfect model and we obtain a $\chi^2/N_{\rm pix} \sim 0.3$ in the absence of instrumental noise. The scatter of the source flux within a list could thus be seen as an additional ``modelling noise". The results for 250\,$\mu$m are shown in Figure \ref{fig:pois}, and we obtain the correct (within 2.5 per cent) total flux density for galaxies as function of magnitude. The total estimate for the CIB is correct to within 1 per cent. This results shows that our fitting routine works well in the absence of correlated sources. 

\begin{figure*}
	\centering  % this centres figure in column
	\includegraphics[trim = 0mm 4mm 0mm 0mm,width = 2.0\columnwidth]{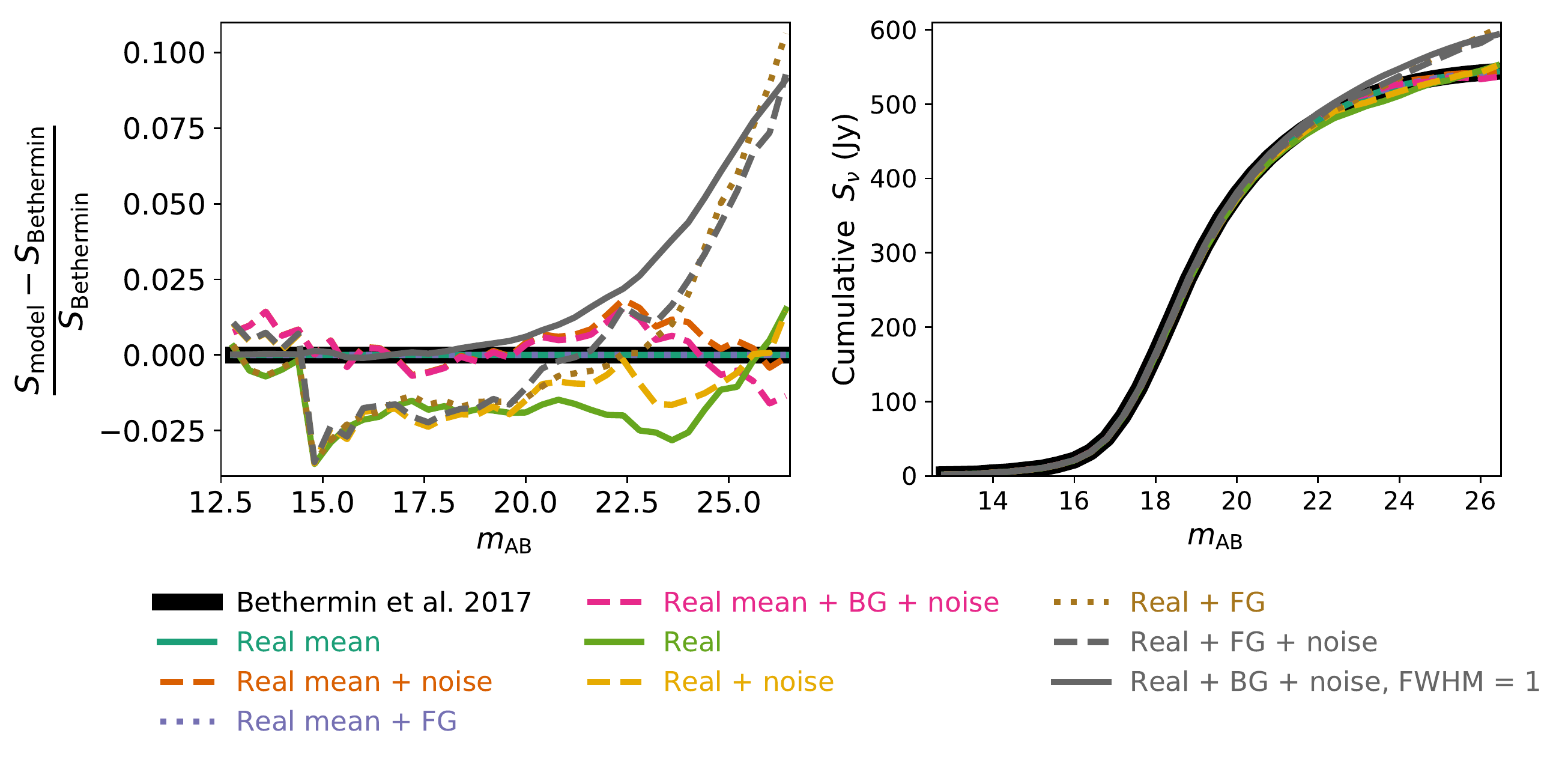}
	\caption{Testing our map fitting method at 250 $\mu$m for clustered sources. The labels are the same as in Figure \ref{fig:pois}, but in all these test the sources have the real (correlated) positions. We overestimate the flux density for faint sources when we allow the foreground to vary. This overestimation also occurs when we create the map with a very small beam (FWHM = 1 arcsec).} 
	\label{fig:real}
\end{figure*}

For the second series of tests we use the actual positions of the sources from the simulation, which means that the galaxies in different lists are correlated. Otherwise, we run the same set of tests as in the previous series. We are able to correctly probe the mean flux densities of galaxy populations, but with two important exceptions (see Figure \ref{fig:real}). These cases are where we overestimate the flux density of faint ($m_{\rm AB}$ $> 20$) galaxies when we allow the foreground to vary while using the individual source flux densities, both with and without noise.\footnote{We performed another test in this series using a FWHM of 1~arcsec to create the map, so that only galaxies within the same pixel are likely to bias each others flux densities. In this case we still obtain the same overestimate as in the nominal resolution (17.6 arcsec) map.}

With a fixed foreground we do not obtain this overestimate. In this case there is a finite amount of flux density available in the map and we cannot interchange flux between galaxies and a foreground. However for the real maps we do not know the value for this foreground and we have to fit for it (while we can set it to zero for the simulations). This overestimate when we fit the foreground simultaneously is potentially worrying, as it could cause an overestimate of the CIB in the real observations. 

The overestimate is primarily caused by very faint sources. We therefore perform a test where we add another three layers of faint sources, between a magnitude of 26.4 and 27.6. These additional 170~000 galaxies contribute only about 0.5 per cent to the CIB. The results for this run are shown in Figure~\ref{fig:more}. This new model leads to an even larger overestimate (13 per cent) of the CIB in the simulations. 

\subsubsection{An over-fitting problem}

A potential cause for this overestimate is ``over-fitting", where the faint sources fit the noise, instead of being assigned a low flux. The results from our FWHM = 1 arcsec test show that this is primarily caused by brighter galaxies in the same pixel.

We ran another test where we created a new map where we add 0.3\,mJy at $250$\,$\mu$m to all faint galaxies with a $m_{\rm AB}> 23.2$ to see if this over-fitting effect is flux dependant. With this simulation the overestimate reduces to $\sim$1 per cent.  For this test the fit of galaxies that are located father away from another galaxy will dominate over this flux exchange between nearby sources on the sky. This flux exchange between galaxies and the foreground remains when we bin our galaxies randomly instead of binning the galaxies according to their magnitude. 

\begin{figure*}
	\centering  % this centres figure in column
	\includegraphics[trim = 0mm 4mm 0mm 0mm,width = 2.0\columnwidth]{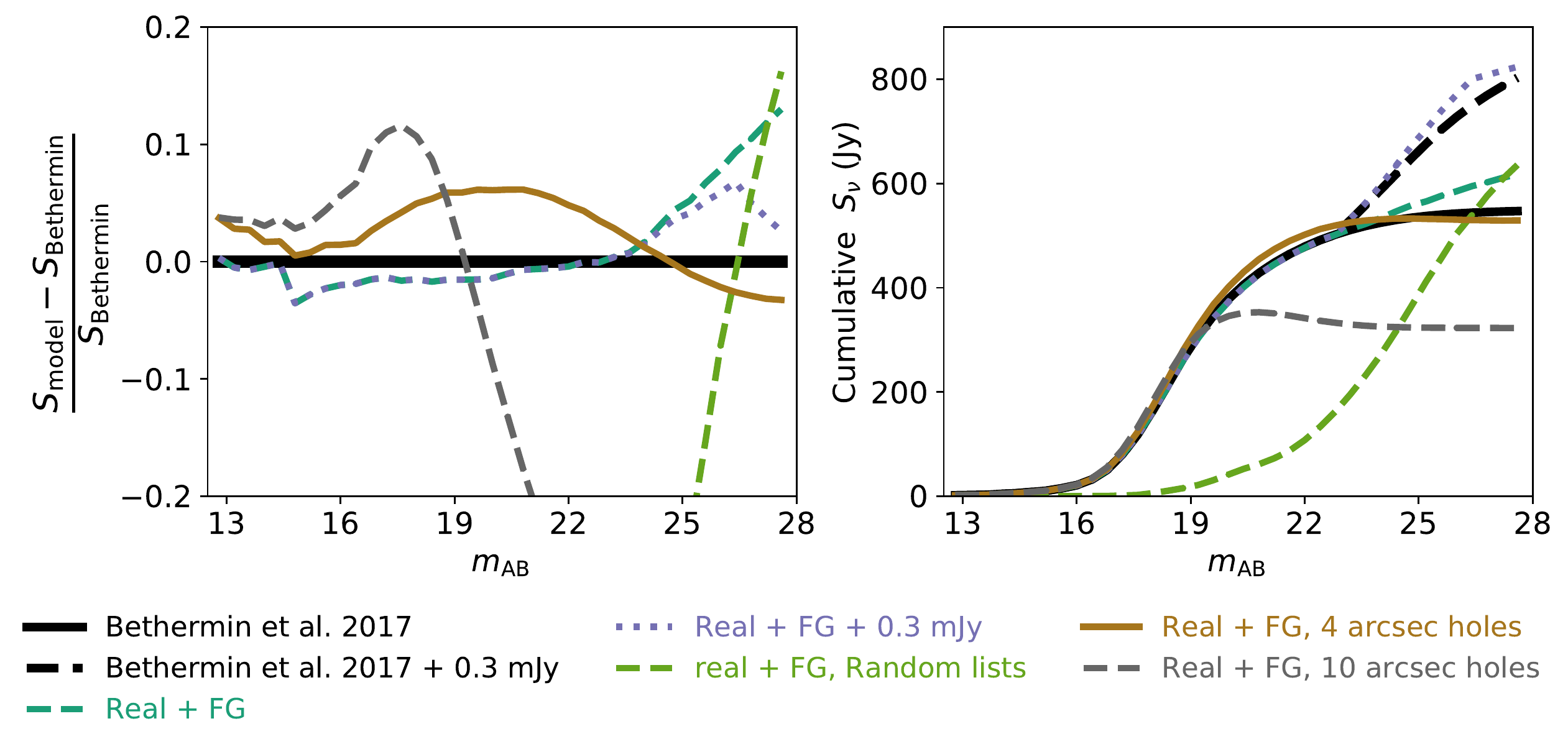}
	\caption{Testing our map fitting method at 250 $\mu$m for deeper simulated data. In black is the ``truth"  from the simulation. On the left we show the offset from the true answer and on the right the cumulative flux density as function of magnitude.  In all tests the sources have the real (correlated) positions. ``+0.3 mJy" indicates that we add 0.3 mJy to faint sources ($m_{\rm AB}$ > 23.2), ``random lists" indicates that we binned the galaxies randomly, and ``holes" indicate that we removed faint sources within the hole radius from a brighter source.  We overestimate the CIB when we fit for the foreground, but this overestimation is diminished when we add 0.3 mJy to faint sources or when we only allow for a maximum of one galaxy within a 4 arcsec radius (removing the  faintest galaxy in a galaxy pair). When we bin our galaxies randomly we obtain the same estimate for the CIB when we bin the galaxies according to magnitude.} 
	\label{fig:more}
\end{figure*}

We perform a test to see if we can eliminate this over-fitting effect by removing the faintest galaxy in every galaxy pair. Where a pair is defined as sources which are within a 4 arcsec radius (when there are multiple matches, than all but the brightest source is removed). With this test we obtain the correct estimate for the CIB (Figure \ref{fig:more}). By removing these sources we obtain a more realistic comparison with medium resolution data, where we would not find multiple sources within a few arcsec due to resolution effects. However, when we make this radius too large then we will \textit{under}estimate the CIB due to the missing sources; we show this by removing all sources within 10 arcsec of a brighter source (Figure \ref{fig:more}). The removal of sources on the sub-arcsecond scale removes both random line-of-sight alignments and galaxies that are located very near each other and are undergoing a merger; these types of sources might be observed as one in the real observations, making this potential over-fitting less of a problem for the real observations. 
     
This overestimate can be explained as follows. Correlated galaxies are more likely to appear near each other on the sky. As both populations of galaxies are fitted simultaneously with our code this should not be a problem. However, if a galaxy population ($A$) is correlated with a population ($B$) {\em and} this correlation is enhanced around bright galaxies from population $A$, then galaxies from $B$ can be assigned the residual (positive) flux density from $A$.

\begin{figure}
	\centering  % this centres figure in column
	\includegraphics[trim = 0mm 0mm 0mm 0mm,width = 1.0\columnwidth]{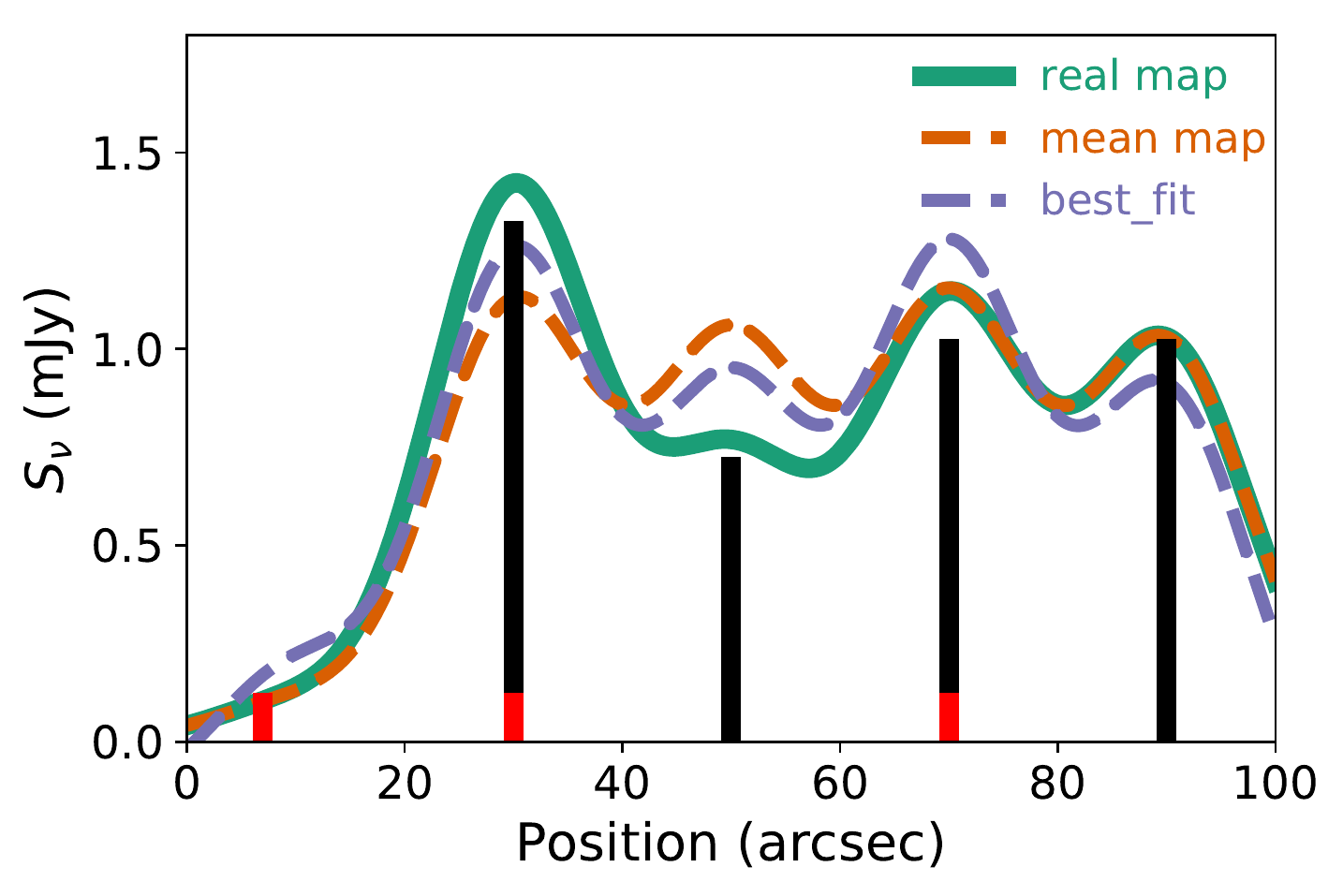}
	\caption{An example of an overestimation of the flux density in a 1-D 250\,$\mu$m map (green). This map contains two populations of sources ($A$, black and $B$, red). The sources in layer $B$ are faint (0.1 mJy) and correlated with the sources in list $A$, with a higher correlation for bright $A$ sources. The purple line shows the best fit of our model and the orange line contains the mean flux of the populations, the result we are looking for.  } 
	\label{fig:cart}
\end{figure}

We can illustrate our explanation in a simpler form (see Figure \ref{fig:cart}).  We make a map containing four sources in layer $A$ and add three correlated sources in layer $B$. We assume we can always obtain an optimal estimate of the mean flux of sources in $A$ (e.g. because they are significantly brighter or more numerous than the $B$ sources).  The four $A$ ($A_1, A_2, A_3$ and $A_4$) galaxies have flux densities of 1.3, 0.7,1.0 and 1.0\,mJy, respectively, with a mean of 1 mJy. Since we have the optimal mean then we have residuals of 0.3, $-$0.3, 0.0 and 0.0 mJy at the four positions of $A$ in the map. The mean of the residuals is zero and the foreground fit will therefore be zero,   the correct answer.  We now add the three correlated $B$ sources  (all 0.1 mJy) at the location of the sources $A_1$ and $A_3$ and one at a random position. After subtracting the (optimal) mean of $A$, the residual flux densities in the map at the position of the $B$ sources are 0.1, 0.4 and 0.1 mJy. The $B$ layer will fit for the mean and obtain 0.2 mJy as an average. The total residuals, after subtracting layer $B$, for our five source locations is $-$0.3\,mJy. This results in a negative foreground fit (Figure \ref{fig:cart}).  It is important to note that this over-fitting would not happen if $B$ were equally correlated with faint and bright $A$ sources. This over-fitting is also reduced if there are many locations of $B$ sources that are not near an $A$ source, as the fit to $B$ will be dominated by the uncorrelated sources.   

With this example we showed an effect not previously accounted for in stacking. Where a bright population ($A$) could cause an overestimate of a faint population ($B$) if the galaxies in $B$ are correlated with the brighter part of the galaxies of sample $A$.

\subsubsection{Comparison with \texttt{SIMSTACK}} \label{sec:ssc}
	
The real observations have masked areas on the sky and we simulate this by: (a) removing the outermost 8 arcsec from the three simulated SPIRE maps; (b) by removing 30 arcsec radius circles around all 392 MIPS sources with $m_{\rm AB}<16$;  and (c) by removing 392 random 30 arcsec radius circles from the map. The bright sources are removed as an examples of saturation by nearby bright galaxies, with the random circles being removed as examples of bright stars, which are not correlated with the galaxies and do not radiate significantly at SPIRE wavelengths.

\begin{figure*}
		\centering  % this centres figure in column
		\includegraphics[trim = 0mm 4mm 0mm 0mm,width = 2.0\columnwidth]{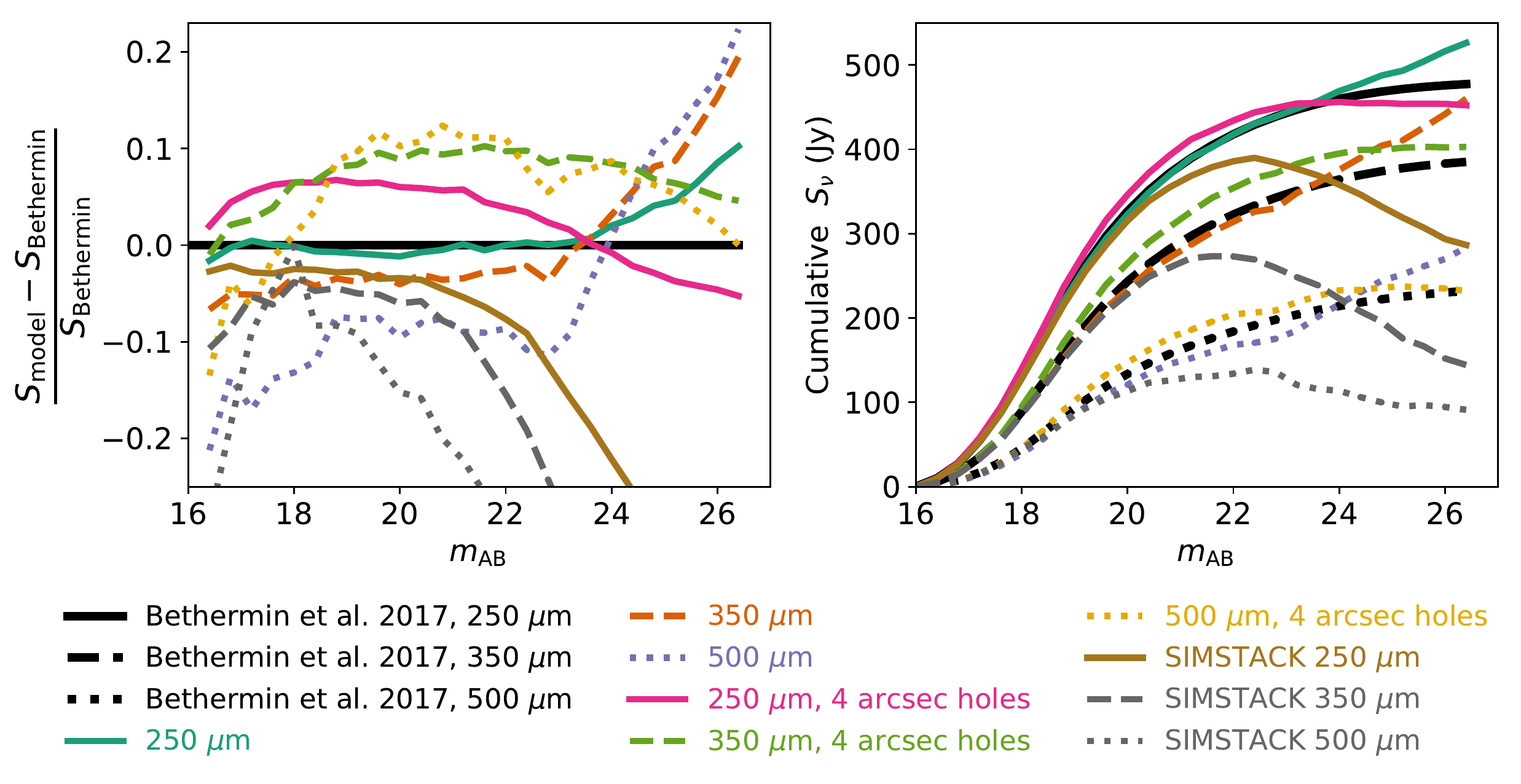}
		\caption{Testing our map fitting method at 250\,(solid), 350\,(dashed) and 500\,$\mu$m \,(dots). In black is the ``truth" from the simulation. On the left we show the offset from the true answer and on the right the cumulative flux density as function of magnitude. ``\texttt{SIMSTACK}" indicates that we did not use our map fitting algorithm, but that we used the basic \texttt{SIMSTACK}. We overestimate the flux density when we fit all the sources in the simulation, due to the effect visualized in Figure \ref{fig:cart}, and we obtain the correct CIB (within 5 per cent) when we remove galaxies within 4\,arcsec of a brighter galaxy. Traditional \texttt{SIMSTACK} underestimates the CIB substantially with $\sim$50 per cent when (almost) all sources in the map are fitted simultaneously.} 
		\label{fig:final_test}
\end{figure*}

All sources within those masked areas are removed from our prior list, and we do not fit the map at those positions. After the removal of masked areas we mean-subtract the map. Due to the large SPIRE beam, there is still flux from sources in the masked areas within the fitted regions of the map. We fit for this flux by adding one extra layer, being the convolution of all the masked  pixels with the SPIRE beam. We now have a simulated map that incorporates instrumental noise and correlated confusion noise, and our prior catalogues contain selection effects from saturation by stars (the random circles) and from nearby bright galaxies (circles around bright sources). We test our algorithm  at all three SPIRE wavelengths in Figure \ref{fig:final_test} and we compare our results with the basic \texttt{SIMSTACK} results.

Our method  outperforms traditional \texttt{SIMSTACK} when measuring the total CIB. When we remove the faint galaxy for galaxy pairs within 4 arcsec (removing the over-fitting effect) we obtain the correct CIB within 5 per cent, where the traditional \texttt{SIMSTACK} method underestimates the total CIB by $\sim$50 per cent (when all galaxies are stacked simultaneously). This underestimation is mainly due to the negative flux density assigned to faint sources ($m_{\rm AB}>22$). 

We over-estimate the CIB by 10-20 per cent when we stack all the galaxies due to the over-fitting effect. In practise these very faint galaxies (with close to zero contribution to the CIB) will not be in our prior catalogue, and this effect can be corrected for by removing the faint galaxy for galaxy pairs within 4 arcsec.

\begin{figure*}
		\centering  % this centres figure in column
		\includegraphics[trim = 0mm 4mm 0mm 0mm,width = 2.0\columnwidth]{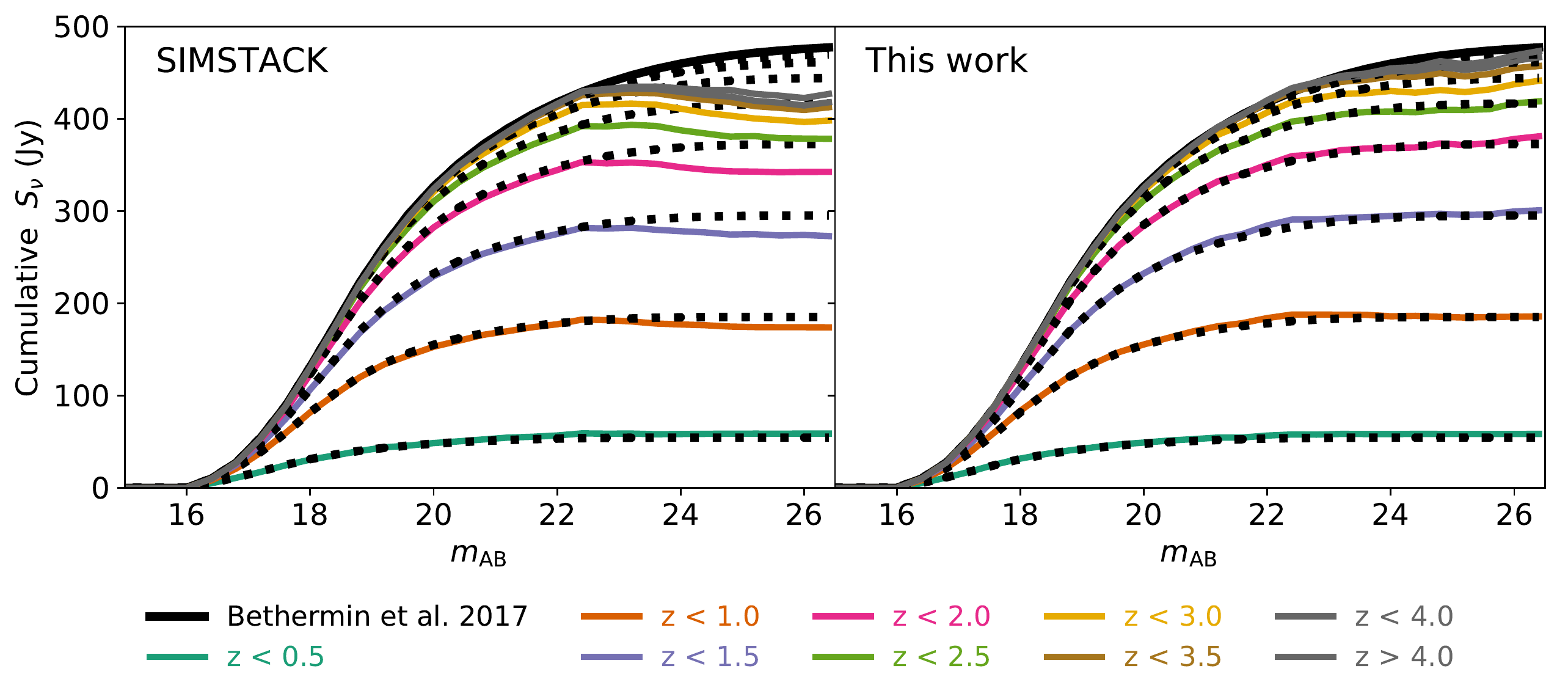}
		\caption{Comparison between SIMSTACK and our map fitting method when we use the redshift of the galaxies. In solid black is the ``truth" from the simulation, and the black dots show the true answer for each redshift range. On the left we fit the redshift slices with \texttt{SIMSTACK}, which underpredicts the flux density for faint galaxies. On the right we use our new map fitting algorithm.} 
		\label{fig:slice}
\end{figure*}

Most papers using \texttt{SIMSTACK} bin the galaxies according to redshift. We test our code and \texttt{SIMSTACK} in Figure\,\ref{fig:slice} with this redshift slicing, where we fit the redshift slices separately from each other. We can see that our code performs very well for galaxies within a $\Delta z = 0.5$ redshift slice, where \texttt{SIMSTACK} underestimates the total CIB by a maximum of 10 per cent. This underestimation only arises when very faint sources ($m_{\rm AB}>23$) are fitted, which are normally not present in the prior catalogues. This suggests that previous results from \texttt{SIMSTACK} are not likely to be incorrect, but that our algorithm is required when a very high ($> 90$ per cent) fraction of the CIB is resolved by the prior catalogue.  Our method is able to correctly calculate the CIB within 1 per cent when redshift slicing is used. 
	
\subsubsection{Incompleteness around bright galaxies}

When a prior list is stacked, we find the total flux density of all the prior sources plus that of correlated coincident sources. Stacking should be done on a mean-subtracted map, so the mean flux from random alignments is zero, leading to a total stacking signal equal to the total flux of the prior sample.  However, when the stacking sample is incomplete for faint objects that are coincident (but not necessarily correlated) with bright objects, a bias occurs. This results in there being a lower probability of finding a randomly aligned bright source at the location of the stacking sample, leading to a foreground of the stack that is lower than the average foreground of the field. For a mean-subtracted map this lower-than-average foreground will be negative, leading to a underestimate for the stacking signal. If the total flux density from the stacked galaxies is less than this negative foreground, then a negative stacking signal can be measured \citep{2013MNRAS.429.1113H}.

We do not see this effect when we fit all sources (Figure \ref{fig:final_test}, 4\,arcsec faint pair remove model), as in this case all the brighter galaxies are fitted simultaneously, leading to a foreground estimate for which the bright sources are taken into account. When we slice in redshifts these bright foreground galaxies are not fitted simultaneously but are just part of the foreground. And when we do not detect faint sources near them on the sky there is a artificial correlation between faint parts in the foreground and the source layer, leading to an underestimate of the source flux density. When we fit all the galaxies simultaneously we do not have the effect described in \cite{2013MNRAS.429.1113H}. We therefore choose to fit all lists of galaxies simultaneously, even if redshift information is available.

\subsection{Final method} \label{sec:sum}

The step by step description of our map fitting procedure is as follows.
	
\begin{enumerate}
		\item Every prior catalogue is binned by AB magnitude with bins ranging from 12.0 to 26.8, with a bin size of 0.4. 
		\item The sources within a bin are used to construct a synthetic $\delta$-function map (+1 for pixels with a source and 0 at locations where there is no source). These maps are convolved with the SPIRE PSFs\footnote{We use a Gaussian PSF having a FWHM of 17.5, 23.7 and 34.6 arcsec for 250, 350 and 500\, $\mu$m, respectively. These are the same PSFs as \citep{2015ApJ...809L..22V} used for the \texttt{SIMSTACK} paper which used the same maps. We note that a change in PSF of order $\sim$0.5\,arcsec changes the results by $\approx 3$\%.} to produce as a fitting-matrix with dimensions M x N where M is number of pixels in SPIRE map and N is number of bins. 
		\item We use the mask provided along with the prior catalogues to re-create the SPIRE maps, with holes at locations where the prior catalogue does not have good data.  
		\item We add two extra layers to our fitting-matrix: one layer models the foreground and is a uniform map i.e. 1 for every pixel; the second layer is the mask\footnote{This mask consist of saturated regions due to both stars and bright galaxies. With more detailed information this method could be improved by using a star mask and a separate bright galaxy mask. } convolved with the SPIRE PSF. This second layer fits the leakage of flux from sources into the map from masked regions.  
		\item The fitting-matrix is used to simultaneously fit all layers using our improved version of \texttt{SIMSTACK}. The layers are fit on all three SPIRE maps independently using Equation \ref{eq:point2}.
		\item We re-run our map fitting algorithm five times with a different bootstrap sample to calculate the errors by calculating the standard deviation from these five measurements. These bootstrap samples come from random re-sampling of the pixels in the map which we use for the fit. 
		\item We re-run our map fitting code 4 times on the map where every time a different quartile is removed. We calculate the effect of sample variance (hereinafter referred to as cosmic variance, for historical reasons) by using these four Jackknife (JK) samples.  		
		\item The mean flux density per magnitude bin is multiplied by the number of sources within the bin to obtain the cumulative flux density as a function of the prior source magnitude (i.e. there is no incompleteness correction).
		\item We calculate the error bars as the quadratic sum of JK errors, bootstrap errors and the SPIRE calibration uncertainty. 
		\item We make another run with our code, where we remove the faint galaxy for every galaxy pair (within 4\,arcsec) to estimate the effects of potential over-fitting, as described in Section \ref{sec:test}.
		\item We use the flux densities derived from the main run (viii) with the error bars calculated in step (ix) to define our upper limit; for the lower limit we use the result from the 4\,arcsec holes run (x), minus our error bar (ix). We than convert the flux density to a surface brightness.
\end{enumerate}

\subsection{Limitations}

The bootstrap error (step vi) gives an estimation of the fitting error, not for the full cosmic variance, as we are still fitting the same sources. The effect of cosmic variance is measured by the JK samples in step vii. We note that the effect of cosmic variance is only measured within the scale of the map, larger scale cosmic variance ($\gg 2\,$deg$^2$ for COSMOS) is not probed by this measurement. We note that the JK errors and bootstrap error are not fully independent, and therefore the quadratic addition sum of the errors is a (small) overestimate.

We cannot formally exclude the possibility that we are over-fitting our real maps in the same way that we over-fit the SIDES simulation. However the maximum source density we fit on the real SPIRE maps is 250\,000\,deg$^{-2}$, while we fit 500\,000\,deg$^{-2}$ for our simulated maps. The over-fitting only affects the faintest of those simulated galaxies, which are (potentially) not detected in the real surveys. 

The problem of over-fitting only arises if faint galaxies are not only correlated with brighter galaxies (brighter in the flux density of the prior catalogue), but also have a higher correlation with the bright end (in the SPIRE map) than with sources that are fainter in the SPIRE maps. An example would be a merger that enhances star formation and therefore SPIRE flux. To determine the magnitude of this effect we need to know the real SPIRE flux densities of the sources, which is what we are trying to find. We do, however, believe that the effect will be smaller than in the SIDES simulation, due to the lower number counts and incompleteness of faint companion galaxies near bright galaxies in the real data. For the SIDES simulation the over-fitting effects cancel out when we remove all faint sources within 4 arcsec of a brighter source. We therefore performed the additional fit (step x) where we remove faint sources in a similar way to obtain a conservative underestimate of the flux density contained in those sources. 

We expect our map fitting estimates to be correct to within 5 per cent, as shown in Section \ref{sec:ssc}. This is comparable with the SPIRE calibration uncertainty and the uncertainty calculated from the JK maps.

Information is lost due to the pixelization of both the SPIRE map and the quantisation of catalogue positions in our source layer (step ii, Section \ref{sec:sum}). The pixelization of the map provides a broadening of the intrinsic telescope beam and thus any fitting will not be as good as it can be.  However, the SPIRE beam size takes this map pixelisation into account and so this does not bias our results. The quantisation of the source positions means that the model beam in the source layer is slightly offset. In the absence of correlations this is effectively broadening the beam (and will bias fluxes low if not taken into account). In the presence of correlated sources this is more complex. In practice we expect these to be very small effects due to the large size of the SPIRE beam compared to the 4 arcsec pixels. The standard deviation of a Gaussian beam profile with FWHM=17.5\,arcsec is 7.4\,arcsec, while the standard deviation of a top-hat response 4\,arcsec pixel is 1.2\,arcsec.  Adding in quadrature we would estimate the additional blurring would produce a beam with standard deviation of $<7.6$\,arcsec.

In future studies, especially with very deep maps (with many scans), maps with smaller pixels can be created to some benefit.  In addition the delta function map can be created with a higher resolution than the map to minimise the impact of the second effect.

\section{Results} \label{sec:res}

\begin{figure*}
	\centering  % this centres figure in column
	\includegraphics[trim = 0mm 0mm 0mm 0mm,width = 2.1\columnwidth]{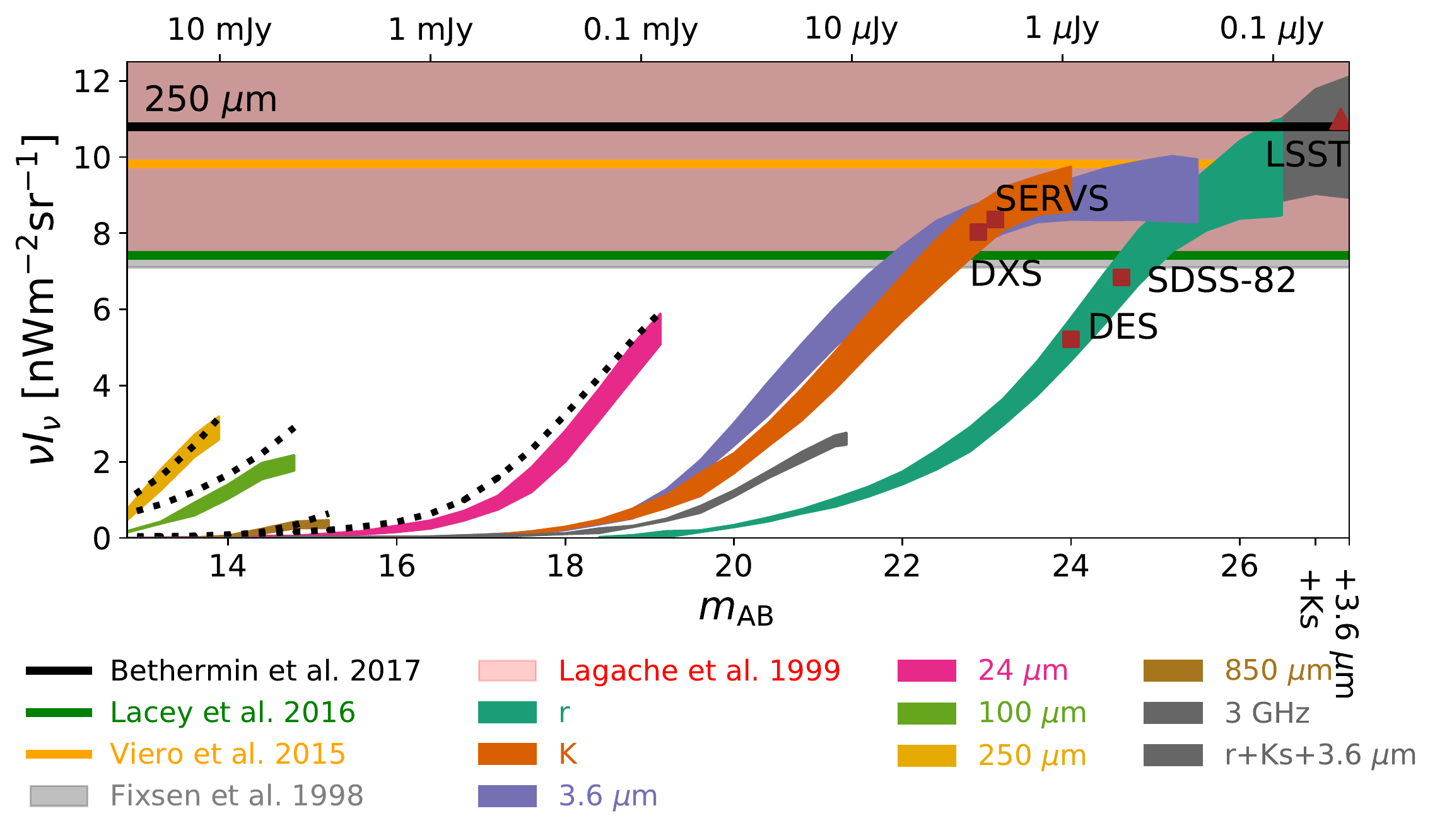}
	\caption{Cumulative measured CIB at 250\,$\mu$m as a function of prior source apparent AB magnitude. The curves $r$, $K_{\rm s}$, 3.6 $\mu$m, 24 $\mu$m, 100 $\mu$m, 250 $\mu$m, 850 $\mu$m and 3 GHz are the estimates from our map fitting with the respective catalogues described in section \ref{sec:prior}. Brown squares show the depth of several current and future large area surveys, with the solid lines show the total CIB as calculated from simulations or previous measurements with SPIRE. The grey and pink shaded areas show the CIB ($\pm 1\sigma$) estimated using FIRAS. The black dotted lines contain the estimates for the CIB from the SIDES simulation contained within FIR prior catalogues. For the $r$-band catalogue we add the 5$\sigma$ $K_{\rm s}$-band and 3.6 $\mu$m detected sources as two extra layers to obtain an estimate for the total CIB. } 
	\label{fig:r250}
\end{figure*}

The results of our map fitting method for 250\,$\mu$m are shown in Figure~\ref{fig:r250}. The best prior catalogues, which reach the highest fraction of the CIB, are the deep optical/NIR surveys. In all these three bands we reach a cumulative flux density that is higher than the 1$\sigma$ lower bounds of the CIB measured by \cite{1998ApJ...508..123F}. With the deep optical/NIR data sets we obtain a very high fraction of the CIB, with our $r$-band stack resolving 9.7 $\pm$ 1.3\,nW m$^{-2}$ sr$^{-1}$ (at $m_{\rm AB}$ = 26.5) which is consistent with FIRAS. We add two more source layers to the $r$-band data, one using the positions of the 5$\sigma$ detected $K_{\rm s}$-band galaxies which are not detected in the $r$-band and the other layer with 3.6\,$\mu$m detected sources which are not detected in the $r$-band or the $K_{\rm s}$-band. With this combination of very deep $r$-band, $K_{\rm s}$-band and 3.6\,$\mu$m priors we obtain a total CIB estimate of 10.5 $\pm$ 1.6\,nW m$^{-2}$ sr$^{-1}$; for this measurements we only use the area (1.38 deg$^2$) with uncorrupted deep $K_{\rm s}$-band data. Our estimates of the CIB are consistent with the total CIB predicted in the SIDES simulation and the total stacked values from \cite{2015ApJ...809L..22V}.

\begin{figure*}
	\centering  % this centres figure in column
	\includegraphics[trim = 0mm 0mm 0mm 0mm,width = 2.1\columnwidth]{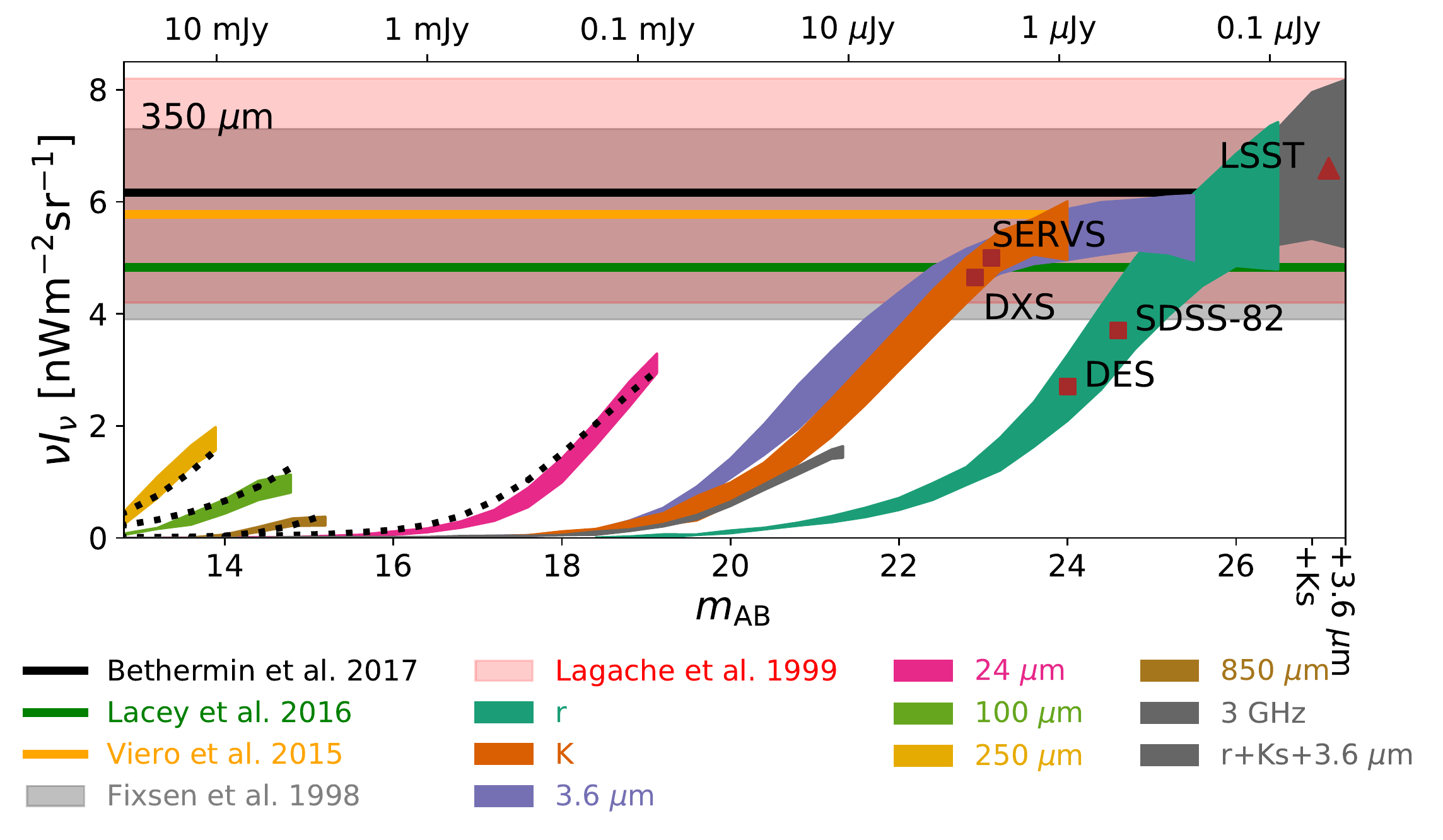}
	\caption{Cumulative measured CIB at 350 $\mu$m as function of prior source AB magnitude. The labels are the same as in Figure \ref{fig:r250}. } 
	\label{fig:r350}
\end{figure*}

\begin{figure*}
	\centering  % this centres figure in column
	\includegraphics[trim = 0mm 0mm 0mm 0mm,width = 2.1\columnwidth]{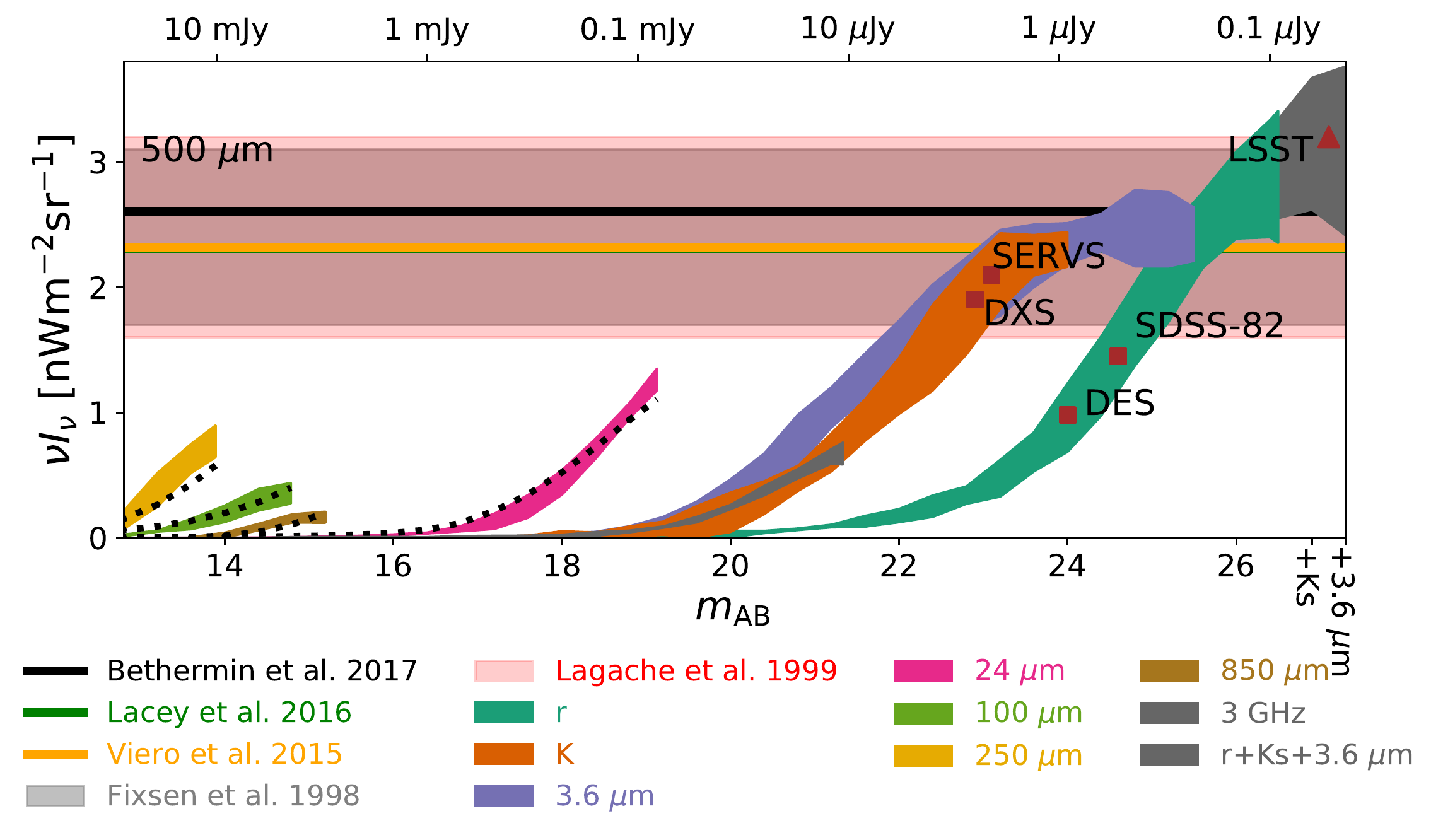}
	\caption{Cumulative measured CIB at 500 $\mu$m as function of prior source AB magnitude. The labels are the same as in Figure \ref{fig:r250}. } 
	\label{fig:r500}
\end{figure*}

The results for 350\,$\mu$m are shown in Figure \ref{fig:r350} and those for 500\,$\mu$m are shown in Figure \ref{fig:r500}. At 350\,$\mu$m we resolve consistent values of the CIB as those measured by FIRAS and \cite{2015ApJ...809L..22V} and simulated by \cite{2017AA...607A..89B}. The total CIB we find is 6.7\,$\pm$\,1.5\,nWm$^{-2}$sr$^{-1}$, with the combination of $r$, $K_{\rm s}$ and 3.6 $\mu$m data. For 500\,$\mu$m we find a total CIB of  3.1\,$\pm$\,0.7\,nWm$^{-2}$sr$^{-1}$, which is higher than (but consistent within 1$\sigma$ with) most previous measurements. The results for all the prior bands can be found in Table \ref{tab:tot}.

\begin{table}
	\begin{tabular}{l c c c} 
		Band & 250\,$\mu$m & 350\,$\mu$m & 500\,$\mu$m \\ \hline
		$r$          & 9.7 $\pm$ 1.3 & 6.1 $\pm$ 1.3 & 2.9 $\pm$ 0.5\\
		$K_{\rm s}$  & 9.2 $\pm$ 0.6 & 5.5 $\pm$ 0.5 & 2.3 $\pm$ 0.1\\
		3.6\,$\mu$m  & 9.1 $\pm$ 0.8 & 5.5 $\pm$ 0.6 & 2.4 $\pm$ 0.2\\
		$24$\,$\mu$m & 5.5 $\pm$ 0.4 & 3.1 $\pm$ 0.2 & 1.3 $\pm$ 0.1\\
		$100$\,$\mu$m& 2.0 $\pm$ 0.2 & 1.0 $\pm$ 0.2 & 0.4 $\pm$ 0.1\\
		$250$\,$\mu$m& 2.9 $\pm$ 0.3 & 1.8 $\pm$ 0.2 & 0.8 $\pm$ 0.1\\
		$850$\,$\mu$m& 0.4 $\pm$ 0.1 & 0.3 $\pm$ 0.1 & 0.2 $\pm$ 0.0\\
		$3$\,GHz     & 2.6 $\pm$ 0.2 & 1.5 $\pm$ 0.1 & 0.7 $\pm$ 0.1\\
		$r$ +$K_{\rm s}$+3.6\,$\mu$m & 10.5 $\pm$ 1.6 & 6.7 $\pm$ 1.5 & 3.1 $\pm$ 0.7
	\end{tabular}
	\caption{The total CIB in units of nW m$^{-2}$ sr$^{-1}$ at the SPIRE wavelengths as measured by our map-fitting algorithm, using different prior prior catalogues.}
	\label{tab:tot}
\end{table}

\begin{figure}
	\centering  % this centres figure in column
	\includegraphics[trim = 0mm 0mm 0mm 0mm,width = 1.0\columnwidth]{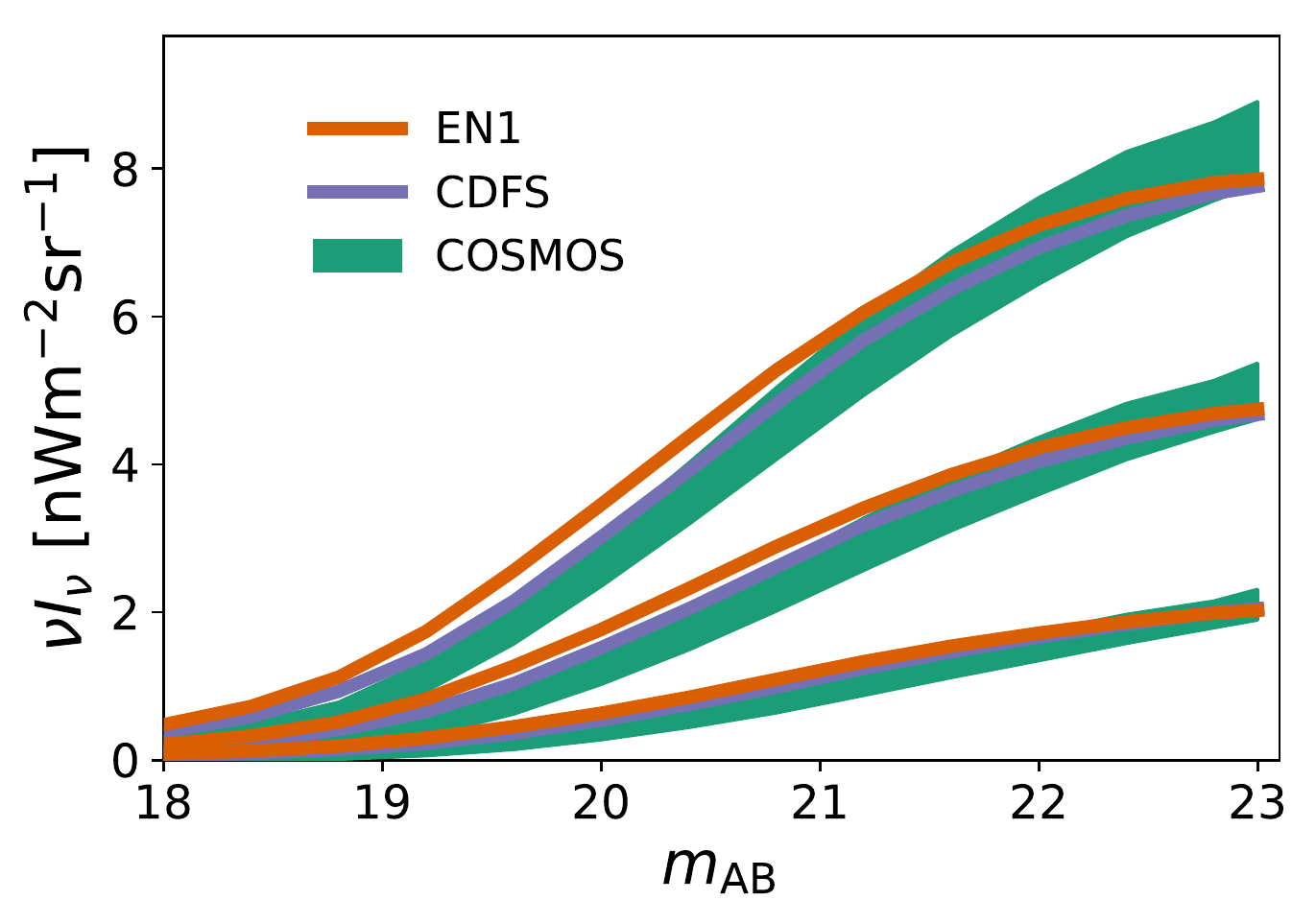}
	\caption{Cumulative CIB at SPIRE wavelengths as a function of IRAC 3.6\,$\mu$m AB magnitude for the EN1, CDFS and COSMOS fields. The top three lines are the measurements at 250\,$\mu$m, the middle three are measured at 350\,$\mu$m and the bottom lines are measured at 500\,$\mu$m.
	We only plot the $\pm 1 \sigma$ error region for the COSMOS field for clarity (the error bars for the other two fields have similar sizes), this error region does not include the JK errors (step vii) as we are now comparing for cosmic variance. The contribution to the CIB from bright galaxies is higher (but not significantly) in the EN1 and CDFS fields.  However, once faint galaxies are included the total contribution to the CIB is higher in the COSMOS field. The differences between the fields is caused by a combination of different masking in IRAC and cosmic variance. } 
	\label{fig:CV}
\end{figure}

We partly calculated the effects of cosmic variance by using our JK samples and our bootstrap error bars. To robustly test the effect of this sampling variance we run our code with IRAC 3.6\,$\mu$m priors on the 2.4\,deg$^2$ ELAIS-N1 (EN1) and the 4.8\,deg$^2$ CDFS-SWIRE (CDFS) field. We also re-run the code for the COSMOS IRAC data, where we make a cut at $m_{\rm AB} =$\,23.1 for all three fields, so that the three fields have  similar depths. The results are shown in Figure~\ref{fig:CV}.   

The difference between the three fields lies mainly in the masking of the IRAC catalogues. The EN1 and CDFS field use the HELP star masks (HELP masks just define the ``holes" from bright stars, not the artefact regions), while the COSMOS field uses a more detailed mask where bright galaxies are more likely to get masked due to saturation of the very deep data.

\begin{figure}
	\centering  % this centres figure in column
	\includegraphics[trim = 0mm 0mm 0mm 0mm,width = 1.0\columnwidth]{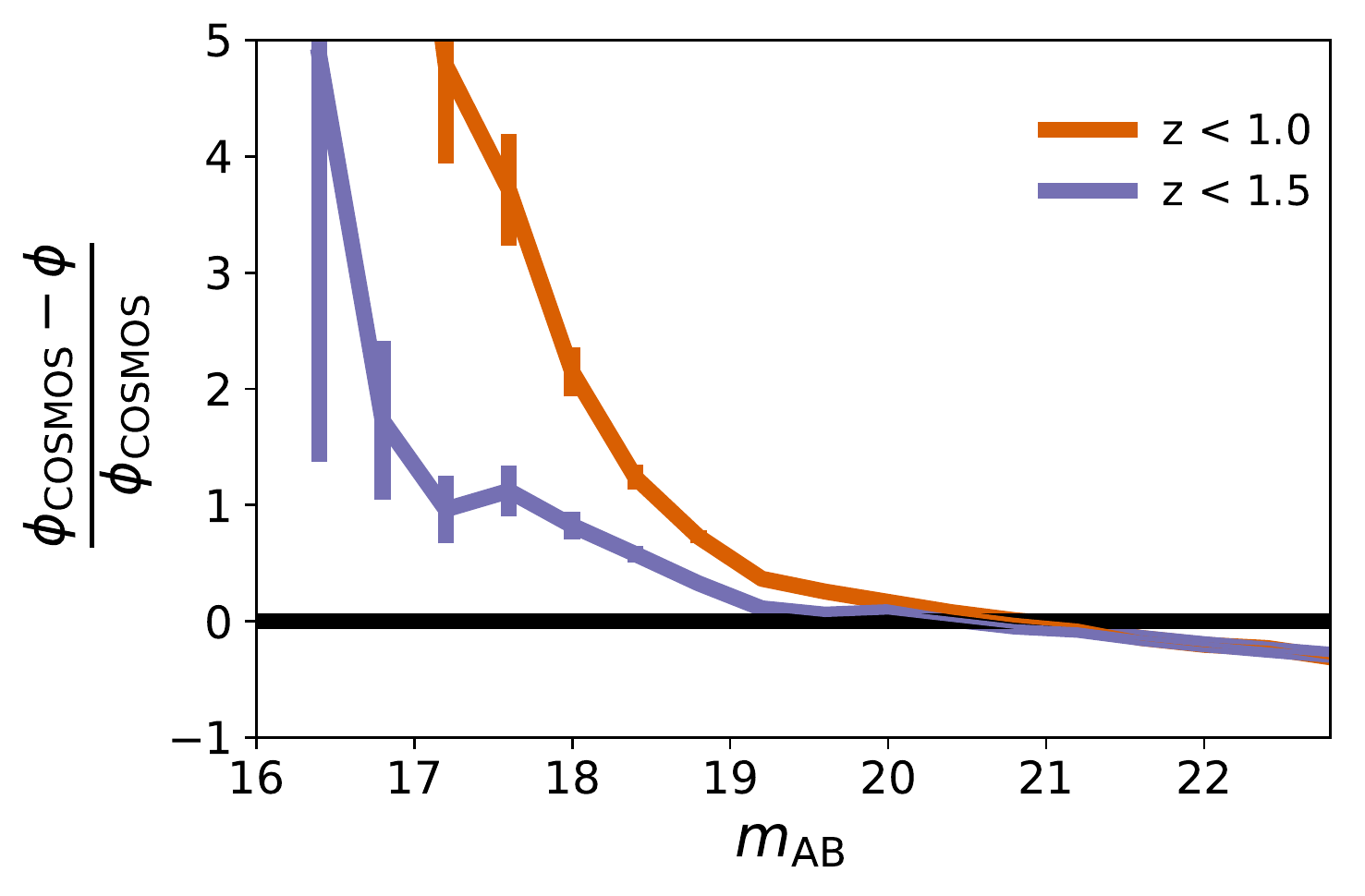}
	\caption{Fractional difference in the number density ($\phi$) of IRAC channel-1-detected objects in the EN1 and CDFS fields with the COSMOS field. Poisson error bars are plotted here. The deep COSMOS field has a lower number density of bright detected objects and a higher number density of faint objects than the larger and shallower fields. } 
	\label{fig:nc}
\end{figure}

The difference in number densities between the three fields is shown in Figure \ref{fig:nc}. It is clear that the number of bright galaxies is much higher in the shallower EN1 and CDFS fields. At the faint end the number of galaxies detected in COSMOS is higher, since it is more complete due to the higher depth. It is also possible that some of the bright objects in the EN1 and CDFS fields are blends of fainter sources, which would have been detected as separate galaxies with the prior-based source extraction code used in COSMOS. This can both explain the excess of bright sources and the lack of faint sources compared to the COSMOS field. These effects of those different number counts can explain the differences in estimated CIB (Figure \ref{fig:CV}).  Even though the measured CIB is different in the three fields they are still consistent within 1$\sigma$ error bars. 

We find that our map-fitting algorithm obtained similar measurement for the contribution to the CIB of catalogued sources in different fields. Our results seem therefore robust against the impacts of cosmic variance.

The results of our code for deeper (and smaller) fields can be found in Appendix \ref{sec:deep}. Those smaller, deeper field (like the CANDELS field) are more prone to cosmic variance, and have larger error-bars due to the smaller sizes. These deep fields are also selected on parts of the sky which avoid bright low redshift galaxies. Which could therefore bias the CIB estimates low.

\section{Discussion} \label{sec:dis}

In Figure~\ref{fig:r250} we also indicate the depth of existing and future large area surveys. Current and ongoing large area $r$-band surveys, such as the 5\,000\,deg$^2$ Dark Energy Survey \citep[DES,][]{2005astro.ph.10346T} and the 300\,deg$2$ SDSS stripe 82 \citep{2014ApJS..213...12J} will detect galaxies responsible for about 50 per cent of the CIB at 250 $\mu$m over these large areas (Figure~\ref{fig:r250}). This area and depth will later be exceeded by the 18,000 deg$^2$ LSST survey \citep{2008arXiv0805.2366I}. The $r$-band depth (27.5) of LSST will be deeper than COSMOS over a huge area and will probe almost all the galaxies responsible for the CIB. It is important to note there likely exists a population of highly obscured (dusty) galaxies at high redshift which even LSST will not see, but will only be visible in small area observations by ALMA and possibly JWST. Wide area $K$-band and IRAC surveys, such as the 35\,deg$^2$ DXS \citep{2007MNRAS.379.1599L} and the 18\,deg$^2$ SERVS survey \citep{2012PASP..124..714M} detect over 75 per cent of the CIB at 250 $\mu$m (Figure~\ref{fig:r250}).

For the total CIB we do not stack on the location of undetected galaxies, which causes an underestimation of the CIB. For galaxies physically nearby our stacked galaxies this may not be a problem, since the flux density will be added to the companion galaxy \citep{2015ApJ...809L..22V}. The missed galaxies are faint at $r$,  $K_{\rm s}$ and 3.6 $\mu$m and are therefore intrinsically very faint or are located at high redshift, which makes it more likely that our 500\,$\mu$m CIB estimate is biased low compared to the shorter wavelength estimates. However, our new determination of the CIB amplitude are higher than most others and provide new lower bounds for the total CIB.

Our CIB estimates are furthermore consistent with the results from \cite{2016ApJ...827..108D}, who calculated deep galaxy number counts at the SPIRE wavelengths using deep priors in the GAMA and COSMOS fields. The obtained number counts were extrapolated to get the number counts for undetected galaxies. The method from \cite{2016ApJ...827..108D} shows an alternative route to use deep prior catalogues to obtain the total value of the CIB, which is corrected for incompleteness, and obtains similar values for the CIB as our measurements.

The absolute FIRAS CIB estimates from \cite{1998ApJ...508..123F} and \cite{1999AA...344..322L} differ by around 10 per cent, and can be considered as an estimate of the systematic uncertainty. These measurements differ in the way the Galactic foreground emission is removed, which provides the main uncertainty in the FIRAS based CIB measurement \citep{1999AA...344..322L}. \textit{Herschel} SPIRE maps have a dramatically better angular resolution than FIRAS (10s of arcsec  vs.  several degrees) and it is therefore possible to remove large scale (few arcmin) Galactic foreground emission. Furthermore, the COSMOS field used in this work lies outside the area of the sky which has high contributions from our own Galaxy. By using the SPIRE data we have removed the largest uncertainty in the CIB measurement. 

The shape of the deep optical and near-infrared lines in Figure \ref{fig:r250}, \ref{fig:r350} and \ref{fig:r500} seem to converge when we go to deeper magnitude ($m_{\rm AB} > 23$). This convergence could potentially be due to incompleteness effects, or it could be that those fainter galaxies have a close to zero contribution to the total CIB. Which raise the interesting prospect that the CIB contribution at $\lambda \le 500\,\mu$m from known galaxies has converged.

\section{Conclusions} \label{sec:con}

In this paper we have developed a novel map fitting algorithm based on \texttt{SIMSTACK} to find the contribution to the CIB from different populations of galaxies. Our code simultaneously stacks all the sources while fitting for the foreground and leakage from masked areas. We tested our code against realistic simulations, which incorporate clustering, confusion noise, instrumental noise and incompleteness effects. Our algorithm outperforms previous stacking algorithms, especially when prior catalogues contain the sources responsible for producing most of the total flux density in the map.
 
We tested our code thoroughly in Section \ref{sec:test}, and our code performs well in confused maps and with prior catalogues that suffer from incompleteness effects.  By testing our method we found a particular kind of bias in stacking/map fitting which can potentially lead to an overestimate of the total value of the CIB. However, these effects are removed in the SIDES simulation by allowing a maximum of one galaxy within a 4\,arcsec radius. We used this approach to recalculate the CIB, finding values that are marginally lower due to the missing sources and the biasing effect. We assume that this effect is smaller in the real data than in the SIDES simulation, since the real data have a lower source density and will miss companion galaxies used to fit the residuals of bright nearby galaxies. Because this effect is smaller in the real data, then our error bars form a conservative lower bound.

We identify a previously unreported bias in stacking/map fitting that could arise when two different lists of prior sources are stacked or fitted simultaneously.  In this case the bright excess of the sources in the first list is fitted by the sources of the second list, leading to an overestimate. This bias is different than the bias discussed in \cite{2013MNRAS.429.1113H}, which is due to incompleteness, and also different from the bias in stacking due to confusion \citep{2013ApJ...779...32V}. 

We used a large range of different prior catalogues in the COSMOS field ($r$, $K_{\rm s}$, 3.6 $\mu$m, 24 $\mu$m, 100 $\mu$m, 250 $\mu$m, 850 $\mu$m and 3 GHz) and divided them up into magnitude bins. Using these bins we measured the total contribution to the CIB as a function of prior source magnitude. We found that compared to the other catalogues the deep ($m_{\rm AB} = 26.5$) $r$-band data resolves the highest fraction of the total CIB at SPIRE wavelengths. 

We add $5\sigma$ detected galaxies in either $K_{\rm s}$ or 3.6 $\mu$m to the $r$-band stack to calculate the total CIB in the maps. Our measurement on the total CIB is 10.5 $\pm$ 1.6, 6.7 $\pm$ 1.5 and 3.1 $\pm$ 0.7\,nWm$^{-2}$sr$^{-1}$ at 250, 350 and 500\,$\mu$m, respectively. The new CIB estimate are consistent with the previous absolute measurements determined using FIRAS data. Our measurements provide new constraints on models that aim to predict the FIR flux from galaxies and can furthermore be used to select the best prior catalogues to deblend the confused SPIRE maps.  Our results show that Future large-area surveys like those with the Large Synoptic Survey Telescope are likely to resolve a substantial fraction of the population responsible for the CIB at 250\,$\mu$m  $\leq \lambda \leq$ 500\,$\mu$m.

\section*{Acknowledgements}
We thank the referee for the useful comments that improved the quality of the work. This project has received funding from the European Union's Horizon 2020 research and innovation programme under grant agreement No. 607254; this publication reflects only the author's view and the European Union is not responsible for any use that may be made of the information contained therein. S.D acknowledges support from the Science and Technology Facilities Council (grant number ST/M503836/1). S. O. acknowledges support from the Science and Technology Facilities Council (grant number ST/L000652/1). 

We would like to thank M.P. Viero for his very useful help with with the SIMSTACK code. 

The \textit{Herschel} spacecraft was designed, built, tested, and launched under a contract to ESA managed by the Herschel/Planck Project team by an industrial consortium under the overall responsibility of the prime contractor Thales Alenia Space (Cannes), and including Astrium (Friedrichshafen) responsible for the payload module and for system testing at spacecraft level, Thales Alenia Space (Turin) responsible for the service module, and Astrium (Toulouse) responsible for the telescope, with in excess of a hundred subcontractors

SPIRE has been developed by a consortium of institutes led by Cardiff University (UK) and including University of Lethbridge (Canada); NAOC (China); CEA, LAM (France); IFSI, University of Padua (Italy); IAC (Spain); Stockholm Observatory (Sweden); Imperial College London, RAL, UCL-MSSL, UKATC, University of Sussex (UK); and Caltech, JPL, NHSC, University of Colorado (USA). This development has been supported by national funding agencies CSA (Canada); NAOC (China); CEA, CNES, CNRS (France); ASI (Italy); MCINN (Spain); SNSB (Sweden); STFC, UKSA (UK); and NASA (USA). 

%%%%%%%%%%%%%%%%%%%%%%%%%%%%%%%%%%%%%%%%%%%%%%%%%%

%%%%%%%%%%%%%%%%%%%% REFERENCES %%%%%%%%%%%%%%%%%%

% The best way to enter references is to use BibTeX:

\bibliographystyle{mnras}
\bibliography{bibli} % if your bibtex file is called example.bib

% Alternatively you could enter them by hand, like this:
% This method is tedious and prone to error if you have lots of references

%\begin{thebibliography}{99}
%\bibitem[\protect\citeauthoryear{Author}{2012}]{Author2012}
%Author A.~N., 2013, Journal of Improbable Astronomy, 1, 1
%\bibitem[\protect\citeauthoryear{Others}{2013}]{Others2013}
%Others S., 2012, Journal of Interesting Stuff, 17, 198
%\end{thebibliography}

%%%%%%%%%%%%%%%%%%%%%%%%%%%%%%%%%%%%%%%%%%%%%%%%%%

%%%%%%%%%%%%%%%%% APPENDICES %%%%%%%%%%%%%%%%%%%%%

\appendix

\section{Deeper fields} \label{sec:deep}

\begin{figure*}
	\centering  % this centres figure in column
	\includegraphics[trim = 0mm 6mm 0mm 0mm,width = 2.0\columnwidth]{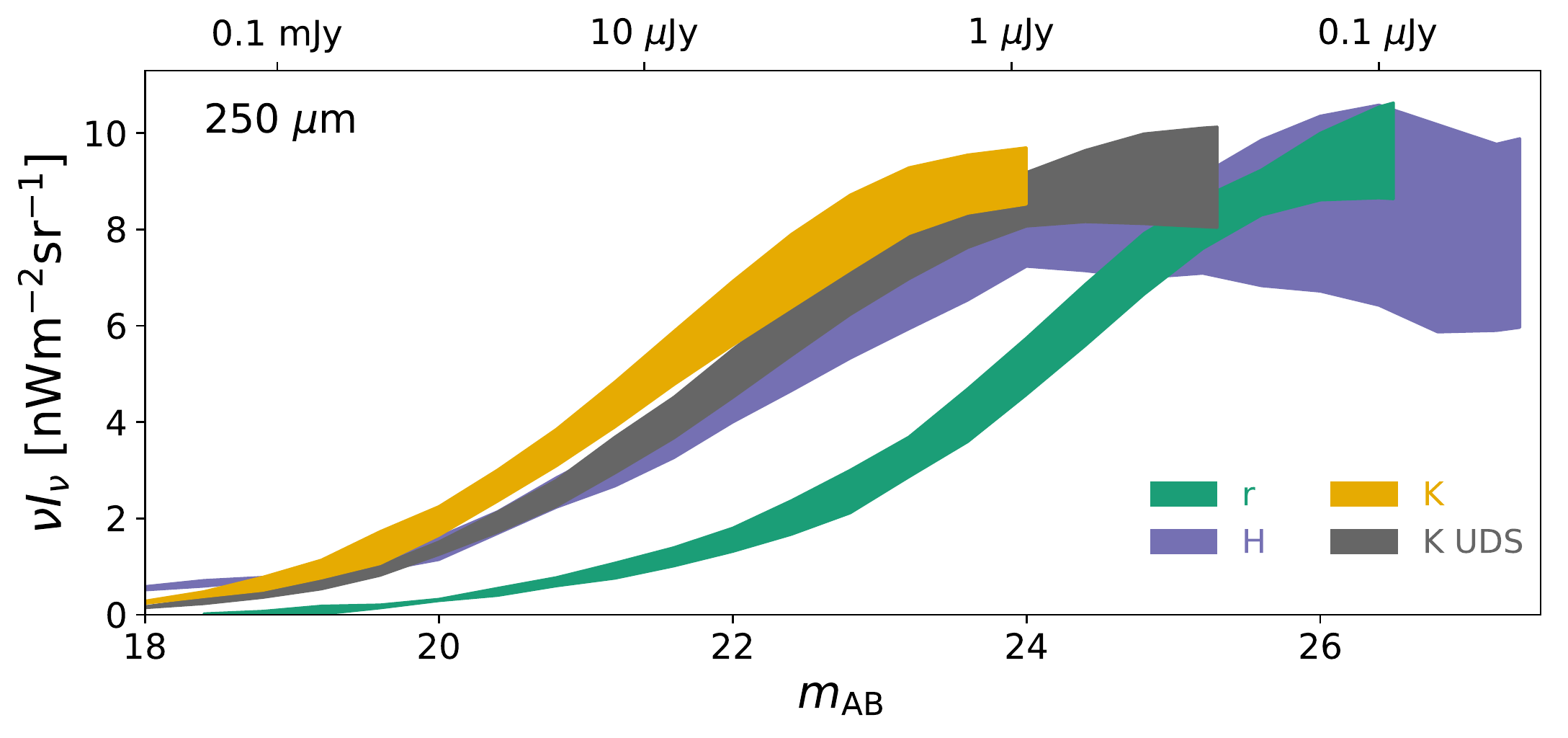}
	\caption{Cumulative measured CIB at 250\,$\mu$m as a function of prior source apparent AB magnitude. The $r$ (green) and $K$-band (yellow) prior catalogues are from the COSMOS field, the $H$-band (purple) catalogue is in the GOODS-S field and the $K$ UDS catalogue (grey) is from the UDS field.   } 
	\label{fig:D250}
\end{figure*}

\begin{figure*}
	\centering  % this centres figure in column
	\includegraphics[trim = 0mm 6mm 0mm 0mm,width = 2.0\columnwidth]{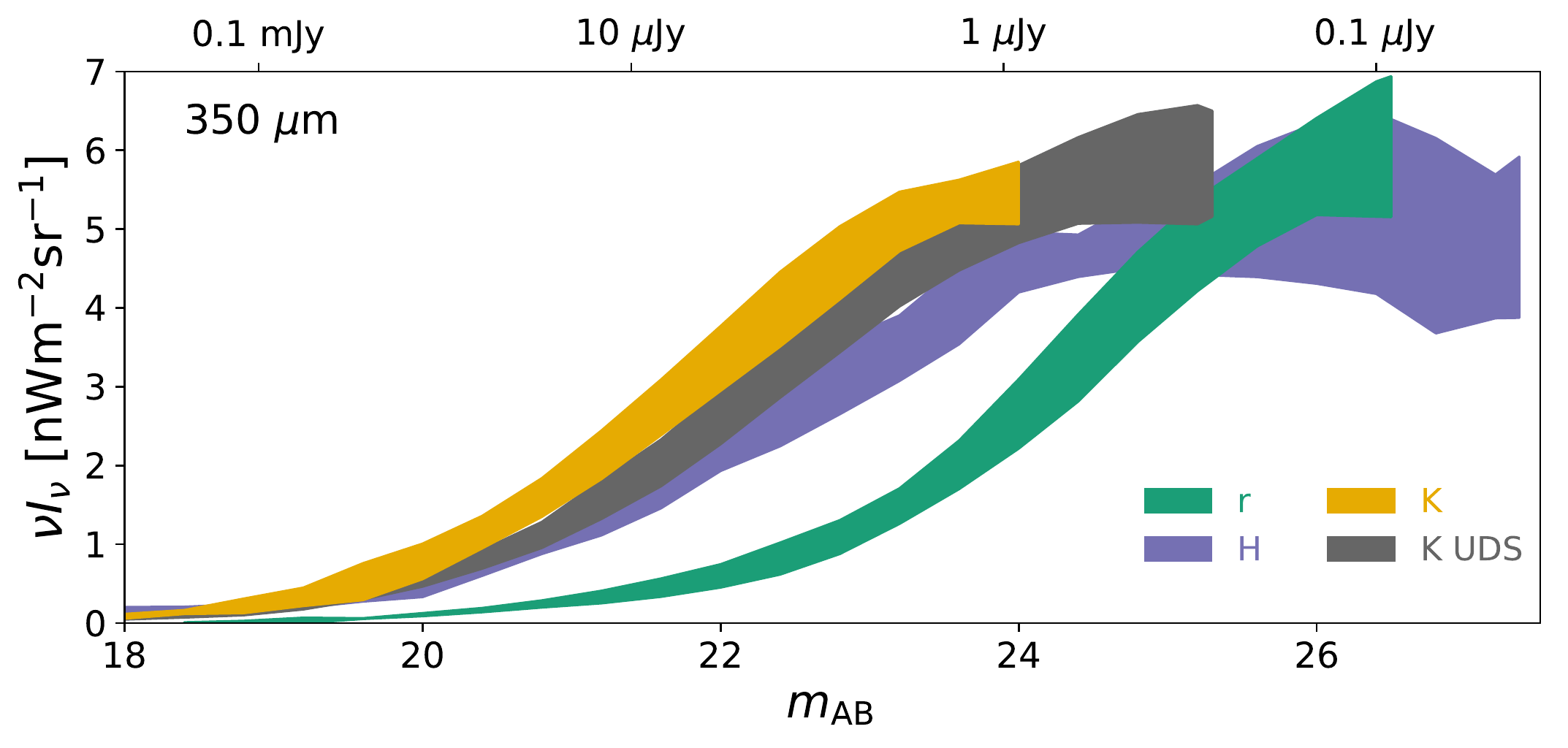}
	\caption{Cumulative measured CIB at 350\,$\mu$m as a function of prior source apparent AB magnitude. The labels are the same as in Figure \ref{fig:D250}. } 
	\label{fig:D350}
\end{figure*}

\begin{figure*}
	\centering  % this centres figure in column
	\includegraphics[trim = 0mm 6mm 0mm 0mm,width = 2.0\columnwidth]{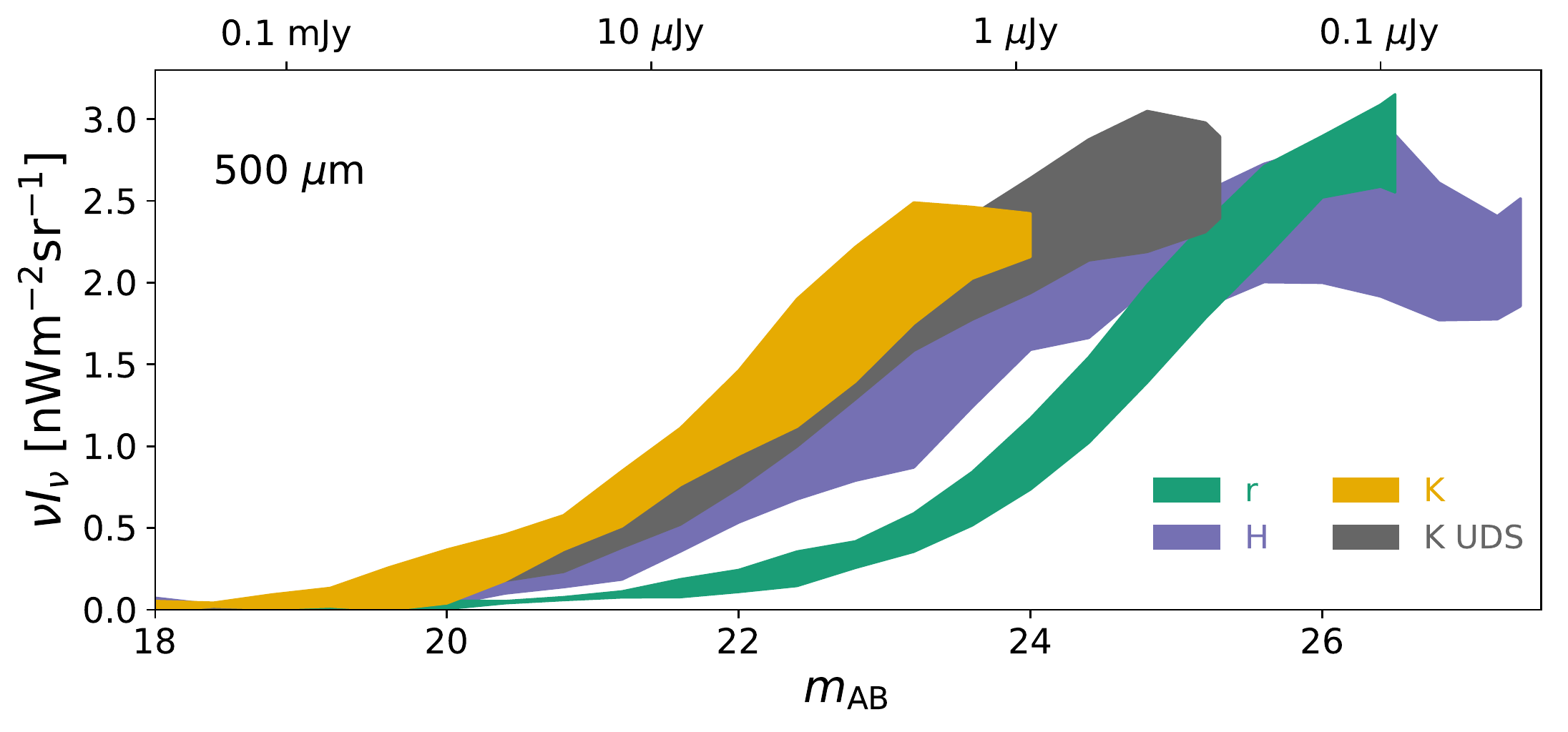}
	\caption{Cumulative measured CIB at 500\,$\mu$m as a function of prior source apparent AB magnitude. The labels are the same as in Figure \ref{fig:D250}. } 
	\label{fig:D500}
\end{figure*}

To find the total CIB for deeper prior catalogues We use the $K$-band from the UKIDSS \citep{2007MNRAS.379.1599L} Ultra-Deep Survey DR11 (UDS, Almaini et al. in preparation), which covers 0.8\,deg$^2$. The $K$-band galaxies are selected up till a depth of 25.3 (AB, $5\sigma$), which is more than a magnitude deeper than our COSMOS run. For the SPIRE maps we use the 3 arcsec pixel maps created for, and used by \cite{2013ApJ...779...32V}. We also use the CANDELS GOOD-S Multiwavelength catalogue \citep{2013ApJS..207...24G} which is selected using the WFC-3 F160W mosaic ($H-$band). The total area covered by this catalogue is only 173 arcmin$^2$. The GOODS-S SPIRE field \citep{2011A&A...533A.119E} is created with 1 arcsec pixels. The $H-$band has a $5\sigma$ limiting depth is 27.36. The comparison with the COSMOS data at 250, 350 and 500\,$\mu$m is show in Figure \ref{fig:D250}, \ref{fig:D350} and \ref{fig:D500}.

With the deeper UDS $K$-band (25.3 $m_{\rm AB}$) data we resolve a comparable fraction of the CIB as with the COSMOS $r$-band (26.5 $m_{\rm AB}$). Compared with the COSMOS $K$-band (24.0 $m_{\rm AB}$) we recover a 0, 7 and 15 per cent higher fraction of the CIB at 250, 350 and 500\,$\mu$m. This bigger difference at longer wavelength indicates that the deeper catalogue detects more galaxies at higher redshift.   

With the very deep $H$-band data from Hubble we do not reach a higher fraction of the CIB compared to the COSMOS $r$-band catalogue. We do however note that the error-bars are larger due to the small size of the field. Our JK error-bars only measure the cosmic variance on similar and smaller scales than the size of the field. For the 173 arcmin$^2$ CANDELS field we are therefore underestimated the total error-bars as we are missing the impact of larger scaler cosmic variance. 

%%%%%%%%%%%%%%%%%%%%%%%%%%%%%%%%%%%%%%%%%%%%%%%%%%

% Don't change these lines
\bsp	% typesetting comment
\label{lastpage}
\end{document}